\documentclass{article}

\usepackage{arxiv}

\usepackage[utf8]{inputenc} 
\usepackage[T1]{fontenc}    
\usepackage{amsmath}
\usepackage{amssymb}
\usepackage{mathtools}
\usepackage{amsfonts}
\usepackage{a4wide}
\usepackage{fancyhdr}
\usepackage{graphicx}
\usepackage[english]{babel}
\usepackage[utf8]{inputenc}
\usepackage[T1]{fontenc}
\usepackage{lmodern}
\usepackage{adjustbox}
\usepackage{booktabs}
\usepackage{xcolor}
\usepackage{array}
\usepackage{multirow}
\usepackage{etoolbox}
\usepackage{graphicx}
\usepackage[bookmarks, colorlinks=true, allcolors=blue]{hyperref}
\usepackage[switch]{lineno}
\usepackage{silence}
\usepackage{siunitx}
\usepackage{subcaption}
\usepackage{tabularx}
\usepackage{tikz}
\usepackage{xcolor}
\usepackage{titlesec}
\usepackage{csquotes}
\usepackage{tikz}
\usepackage{tikz-cd}
\usepackage{bm}
\usepackage{algorithm}
\usepackage{algpseudocode}
\usepackage{url}
\usepackage{enumitem}
\usepackage[nomargin,inline,draft]{fixme}
\usepackage{adjustbox}
\usepackage{amsthm}
\usepackage[backend=bibtex, sorting=none, backref=true,style=numeric-comp]{biblatex}
\usepackage{placeins}
\usepackage[capitalize,noabbrev]{cleveref}
\usepackage{microtype}
\usepackage{pifont}
\usepackage{parskip}

\newcommand{\cmark}{\ding{51}}%
\newcommand{\xmark}{\ding{55}}%
\title{Deep Generative Models for Detector Signature Simulation: A Taxonomic Review}

\author{ 
    \href{https://orcid.org/0000-0003-4095-9657}{\includegraphics[scale=0.06]{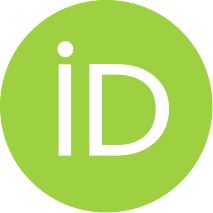}\hspace{1mm}Baran Hashemi~\footnote{The author changed his first name from Hosein to Baran}}\\
	ORIGINS Data Science Lab\\
	Technical University Munich,\\
	Munich, Germany \\
	\texttt{baran.hashemi@origins-cluster.de} \\
	\And
	\href{https://orcid.org/0000-0003-0924-3036}{\includegraphics[scale=0.06]{orcid.pdf}\hspace{1mm}Claudius Krause} \\
	Institute for High-Energy Physics~(HEPHY), \\
    Austrian Academy of Sciences (OeAW),\\ Vienna, Austria \\
	\texttt{Claudius.Krause@oeaw.ac.at} \\
}

\renewcommand{\headeright}{Under Review}
\renewcommand{\undertitle}{Submitted to Reviews in Physics}
\renewcommand{\shorttitle}{Deep Generative Models for Detector Signature Simulation}

\hypersetup{
pdftitle={Deep Generative Models for Detector Signature Simulation: A Taxonomic Review},
pdfauthor={Baran Hashemi, Claudius Krause},
pdfkeywords={First keyword, Second keyword,},
}

\DeclareFieldFormat{urldate}{}
\bibliography{refs}

\begin{document}
\footnote{The author changed his first name from Hosein to Baran}
\maketitle

\begin{abstract}
   In modern collider experiments, the quest to explore fundamental interactions between elementary particles has reached unparalleled levels of precision.
   Signatures from particle physics detectors are low-level objects~(such as energy depositions or tracks) encoding the physics of collisions~(the final state particles of hard scattering interactions). 
   The complete simulation of them in a detector is a computational and storage-intensive task. 
   To address this computational bottleneck in particle physics, alternative approaches have been developed, introducing additional assumptions and trade off accuracy for speed.
   The field has seen a surge in interest in surrogate modeling the detector simulation, fueled by the advancements in deep generative models. These models aim to generate responses that are statistically identical to the observed data.
   In this paper, we conduct a comprehensive and exhaustive taxonomic review of the existing literature on the simulation of detector signatures from both methodological and application-wise perspectives.
   Initially, we formulate the problem of detector signature simulation and discuss its different variations that can be unified. 
   Next, we classify the state-of-the-art methods into five distinct categories based on their underlying model architectures, summarizing their respective generation strategies. 
   Finally, we shed light on the challenges and opportunities that lie ahead in detector signature simulation, setting the stage for future research and development.
\end{abstract}


\section{Introduction}
Simulation is at the core of research in experimental particle physics. It allows for a detailed comparison of experimental data and fundamental theory, and therefore for a proper interpretation of measurements. Following the hierarchy of the involved energy scales, the simulation chain is split into the simulation of the hard~(high-energy) scattering interaction; the decay of the heavy, unstable particles and the evolution of a parton shower; the hadronization of all colored particles to color-neutral ones; and finally, the interaction of these stable particles with the detectors~\cite{butter_machine_2023,campbell_event_2024}. 
In this review, we will focus on the last step, how the signatures of particles with given initial conditions~(like sensor positions, particle type or energy) in the detector look like, as this is usually the computationally most-expensive step. The most detailed and accurate simulation toolboxes at hand for this task are Geant4~\cite{agostinelli_geant4simulation_2003,allison_geant4_2006,noauthor_recent_nodate} and FLUKA~\cite{ferrari_fluka_2005,noauthor_fluka_nodate}. These track each particle and all the secondary particles it produces via fundamental interactions through all different types of matter in the entire available geometry. These simulations result in a wide variety of representations for different types of detector signatures. Among them, the showers of particles in calorimeters; the collimated, narrow cones of hadrons~(called jets) produced during the hadronization of colored partons; and hit patterns that form tracks of charged particles --- all yielding datasets with different granularity, complexity and characteristics. 
 
However, the complexity and accuracy of these simulations come at the price of being computationally expensive, easily becoming the bottleneck in the simulation chain. This led to the development of several faster alternatives, trading accuracy for speed. For example, each experimental collaboration developed their own so-called ``Fast Simulation'', a dedicated detector simulation making several simplifying assumptions on both, geometry and physics, leading to a speed-up of $\mathcal{O}(10)$--$\mathcal{O}(100)$~\cite{abdullin_fast_2011,atlas_collaboration_atlfast3_2022}. If one is only interested in few, high-level observables like isolated leptons or missing transverse energy, one can resort to Delphes~\cite{de_favereau_delphes_2014,selvaggi_delphes_2014, mertens_new_2015}. 
It is extensively used in phenomenological studies, especially also since experiment-specific Fast Simulations are restricted to members of the respective collaborations. By including even the reconstruction steps, one achieves a very fast end-to-end simulator called FlashSim~\cite{vaselli_flashsim_2023}. The common denominator of all of these approaches is that the underlying physics is stochastic: Given the same initial conditions, the resulting recorded patterns in the detector are random samples drawn from a~(complicated) distribution.

This is where Deep Generative Models~(DGMs) will be useful. The core concept of generative models is derived from the training of a density estimator that produces samples approximating the distribution of the training data. 
The essence of generative models is rooted in the domain of density estimation~\cite{parzen_estimation_1962,b_w_silverman_density_1998}, where various methods~\cite{fukunaga_optimization_1973,magdon-ismail_neural_1998,cranmer_kernel_2001}, both parametric and non-parametric approaches, were employed to estimate the underlying distribution of data.
The initial wave of neural network-based generative models, also known as energy-based models~\cite{hinton_optimal_1983,hinton_training_2002}, attempted to accomplish this by establishing an energy function proportional to the likelihood for data points. 
However, these models faced challenges in scaling up to high-dimensional, complex data, like natural images. 
Thus, they necessitated the use of Markov Chain Monte Carlo~(MCMC) sampling~\cite{hinton_fast_2006}, a method required during both training and inference stages. The goal of energy-based models is to adjust the model parameters so that the energy function assigns lower energy~(\textit{i.e.}, higher likelihood) to real data points and higher energy to improbable ones. However, the challenge is that these energy landscapes can be very complex, especially for high-dimensional data. Calculating likelihoods or gradients directly becomes computationally infeasible or extremely difficult. That is where MCMC comes in. This method, characterized by its iterative nature, often resulted in a very slow and inefficient process. 
The availability of larger datasets and significant advancements in DGMs~\cite{tomczak_deep_2022} for natural images led to a resurgence of interest in generative models in the past few years.
These DGMs have pushed the boundaries in terms of visual quality, sample diversity, and speed of sampling. 

In Particle Physics, the application of DGMs was first studied in the ``Fast and Efficient Simulation'' campaign~\cite{de_oliveira_learning_2017} that sparked the search for faster and more storage-efficient surrogate models and simulation methods of collider physics experiments~\cite{adelmann_new_2022,butter_machine_2023}. 
Surrogate models for fast detector simulation are simplified, computationally efficient approximations that emulate the behavior of more complex, detailed simulations of signatures from particle detectors.
These models are constructed using DGMs, trained on a dataset either generated by the Geant4 simulation~\cite{agostinelli_geant4simulation_2003,allison_geant4_2006,noauthor_recent_nodate} or the real signatures from particle detectors. 
Once trained, the model can generate samples that are statistically similar to their training data and even can generalize to the data beyond the training regime~\cite{butter_ganplifying_2021,bieringer_calomplification_2022,matchev_uncertainties_2022,hashemi_deep_2024} depending on the objective. 
There are some key requirements that the model should meet to be effective and efficient:

\begin{enumerate}
    \item \textbf{Low time-complexity}: The surrogate model must be computationally efficient in order to facilitate Fast Simulations. Computational speed is crucial when performing large-scale simulations or when needing to iterate the model many times for optimization or fine-tuning. Thus, the model should take advantage of parallel processing capabilities and have to be optimized for the hardware it is expected to run on, whether it is a CPU or GPU.
    
    \item \textbf{Low space-complexity}: It should come with a minimal storage cost. Hence, the underlying compression technique has to reduce the storage footprint without sacrificing too much of the precision and accuracy of the downstream physics analysis.
    
    \item \textbf{Realistic and Diverse:} It has to generate samples as faithful and diverse as possible from the downstream physics analysis point of view. Thus, the sampling techniques should be capable of employing the nuanced behaviors and symmetries of the detector as an inductive bias to ensure that the model captures the diversity inherent in the real data.
    
    \item \textbf{Generalization:} 
    For certain objectives, it has to be able to extrapolate to parameter levels~(such as energy levels, kinematic profile, beam parameters, and luminosities) beyond the current experimental limits in order to analyze the detector's effect and do physics analysis in the extrapolation region. Therefore, the model should be robust against overfitting and incorporate a proper measure of control when extrapolating to give a range of plausible outcomes.
\end{enumerate}

The subsequent sections of this paper are organized in the following manner. We begin by defining the challenge of simulating signatures from particle detectors with deep generative models as surrogates. Subsequently, we present a taxonomic breakdown of current methodologies, categorizing them into five classes. 
We introduce a universal framework, delve into prevalent generation strategies and tasks, and then provide an exhaustive review of research for each category. 
In our concluding remarks, we address the challenges and potential advancements in deep generative modeling for detector response generation. 
Through this survey, we aim to offer a unified and consolidated view of the present landscape of deep generative models for detector signature simulation. 
We hope to bridge the existing gap in this sparse and occasionally overlooked research, ensuring an inter-experimental perspective that remains objective and free from bias.

\section{Problem Definition}
The detector signatures can be defined by a triplet $\mathcal{D}_{e} = (\mathbf{L}_e, \mathbf{C}_e, \mathbf{H}_e)$ for each event $e$, where $\mathbf{L} \in \mathbb{Z}$ is the detector component such as layers or sensors indicators, $\mathbf{C} \in \mathbb{R}^{n}$ is the global attribute of the corresponding event such as incident energy or the beam parameter, $\mathbf{H} \in \mathbb{R}^{d}$ is the hit points per sensor/layer like particle tracks, energy clusters or jets, which can be represented for instance by a grid (like an image), by a sequence, or by a multi-set (like point clouds and graphs). 
Given a set of $M$ observed detector responses $\mathbf{D} = \{\mathcal{D}^i_e\}_{i=1}^M$, DGMs learn the distribution of these signatures $p(\mathbf{D})$, from which new responses can be sampled $\mathbf{H}_{\text{new}} \sim p(\mathbf{H})$.

In the context of detector simulation for physics analyses, $\mathbf{C}$ would be defined implicitly by the underlying hard scattering process and subsequent shower and decays, yielding a distribution of incident parameters for each particle~(per event) to be simulated. The direction of motion of these particles also implicitly defines which component of the detector $\mathbf{L}$ will be relevant for the simulation. In the context of working with DGMs, these parameters are usually sampled from a convenient, externally-given distribution (for example log-uniform incident energies) or even fixed completely (for example by considering only particles traveling perpendicular to the detector surface). 

This paper presents both latent variable and non-latent variable approaches as the mainstream detector response generation models. Latent variable approaches follow an encoder--sampler--decoder pipeline. It firstly maps the data into a hidden (latent) space through an encoding function, manipulates the hidden variables to reflect the desired properties of the detector response to be generated, and then generates new samples based on latent codes through a decoding function. 
Unlike the latent variable models\footnote{
Here, the term latent refers to a compressed space. In the literature, latent spaces sometimes also refer to uncompressed spaces, such as for normalizing flows or diffusion models.}, non-latent variable approaches directly map the input data to the desired detector responses stochastically without the intermediary step of encoding into a compressed hidden space. 

\section{Algorithm Taxonomy}
Within the scope of latent variable techniques, the given data is mapped into a stochastic latent space. 
An \textit{i.i.d} sample from this distribution is then fed into a decoder that reconstructs the original data structures. In non-latent variable models, generation methods do not map the observation into the latent space. In other words, they perform the generation in the raw space without first compressing it into a latent form and directly generate the data representation. 
These models, given initial conditions, apply transformations or rules that map~(encoder) the original input features straight to the output. Prior to delving into an in-depth exploration of distinct models, we initially established a bird-eye-view pipeline that encompasses encoding, sampling, and decoding stages. This pipeline allows us to encapsulate the majority of the existing generative models used for detector simulation within a single unified framework. In accordance with this framework, we will classify various methodologies based on their interaction with the following three pivotal components:

\textbf{The Encoder}. The encoding function $f_{\theta}(z|\mathcal{D})$ maps the detector response triplets to a dense, continuous~(or quantized) topological space. To ensure the learned latent space is meaningful for generation, depending on the data type and the inductive bias, one incorporates various morphisms~(\textit{e.g.}, Convolutional Neural Networks~\cite{fukushima_neocognitron_1980}, Graph Neural Networks~\cite{velickovic_everything_2023}, DeepSets~\cite{zaheer_deep_2018}, etc.) as the encoder. In this way, the encoder function $f_{\theta}$ generates the parameters for a stochastic distribution that adheres to a prior distribution denoted as $p(z)$.

\textbf{The Sampler}. Following the encoding process, the model that generates the detector response draws samples from the latent distribution $z \sim p(z)$. There are generally two prevalent strategies for this sampling process: \textbf{random sampling} and \textbf{controlled sampling}~\cite{goodman_controlled_1950}. 
Random sampling refers to randomly selecting latent vectors from the learned or prior distribution. 
On the other hand, guided or controlled sampling is designed to sample the stratified latent vectors with the specific goal of generating samples that exhibit certain preferred characteristics. 
These characteristics could be certain kinematic profiles, beam parameters, physics processes, number of hits, luminosities, incident angles, or incident energies. In most tasks, controlled sampling is model-dependent and necessitates an additional optimization component that goes beyond the scope of random generation.

\textbf{The Decoder}. Upon obtaining the latent features drawn from the learned distribution, the decoding mechanism is responsible for
reconstituting them into a data manifold. Due to the multi-objective and fine-grained learning nature of detector simulation, the decoding phase is inherently more complex than the encoding stage.  In other words, it is of paramount importance that the underlying symmetries/inductive biases of the detector signature is satisfied at the decoding stage.
Typically, the decoders could be grouped into three categories: \textbf{sequential generation}, \textbf{one-shot generation}, and \textbf{zero-shot generation}. 
Sequential generation refers to generating the detector component information in a set of consecutive and autoregressive steps, usually done sensor-by-sensor, layer-by-layer, or hit-by-hit. 
One-shot generation, instead, refers to generating the whole detector signatures in one single step. 
Zero-shot generation is when the model generates an unseen set of detector information where the control parameter goes beyond the training data. Thus, the model has to generalize well to the Out-Of-Distribution~(OOD) domain. While zero-shot generation may employ one-shot or sequential methods for the actual data construction, its main challenge lies in ensuring that these generated responses are both accurate and plausible when dealing with conditions that differ from the training data.

\subsection{Deep Generative Models}
\begin{figure}[!htb]
    \centering
    \includegraphics[width=\linewidth]{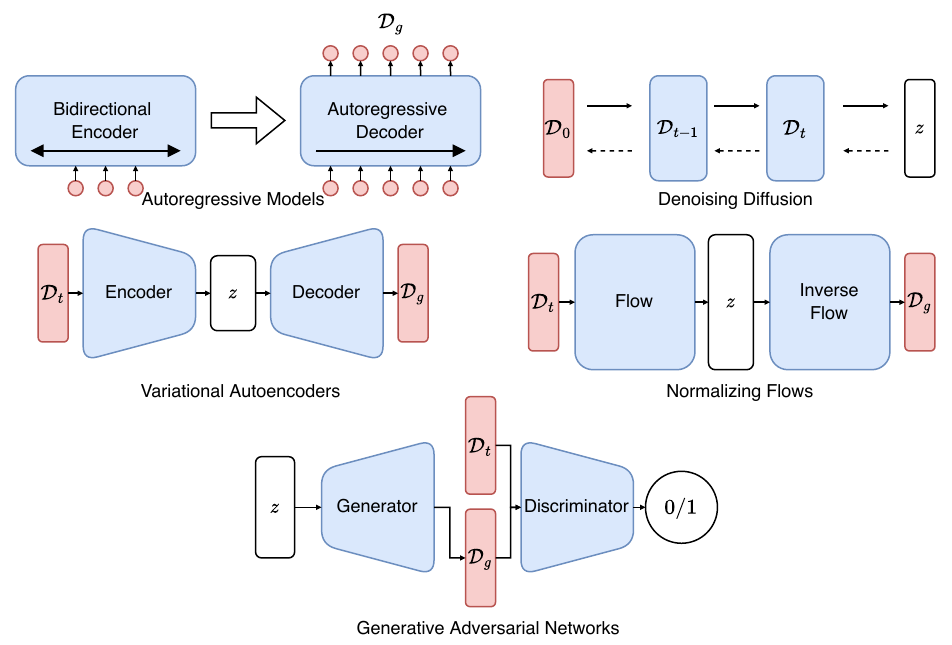}
    \caption{Generic architectures of deep Generative models for simulating detector signatures.}
    \label{fig:generative}
\end{figure}

At first, we discuss the following five representative generative models, depicted in~\cref{fig:generative}, and their capabilities for simulating detector signatures, summarized at~\cref{tab:compare}.

\textbf{Variational Auto Encoders~(VAE).} VAEs learn a lower-dimensional representation of the data in the latent space by forcing the information through a bottleneck.
A VAE estimates the distributions of data $p(\mathcal{D})$ with the amortized variational posterior~\cite{kingma_auto-encoding_2022,rezende_stochastic_2014}, by maximizing the Evidence Lower BOund~(ELBO), \textit{i.e.} the lower bound of the log-likelihood function, as follows:

\begin{equation}
    \mathcal{L}_{\text{VAE}} = \mathbb{E}_{z\sim f_{\theta} (\bm{z} \mid \mathcal{D})} \left[ \log(p_{\phi}(\mathcal{D} \mid \bm{z})) \right] - D_{\text{KL}} \left( f_{\theta}(\bm{z} \mid \mathcal{D}) \parallel p_{\phi}(\bm{z}) \right)
\end{equation}

where the former term is known as the negative reconstruction loss between the input and the reconstructed data, while the latter is the disentanglement enhancement regularizer that drives the amortized variational posterior $f_{\theta}(z|\mathcal{D})$ to the prior~(marginal) distribution $p_{\theta}(z)$.
For simulation-based inference where the tractability of the likelihood is vital, one can use various sampling methods to estimate the likelihood such as Importance Sampling~\cite{rezende_stochastic_2014,burda_importance_2016}, Annealed Importance Sampling~\cite{neal_annealed_2001,zhang_differentiable_2021}, and Nested Variational Inference~\cite{zimmermann_nested_2021} with a sufficiently large sample space to estimate the log-likelihood. 
Generally, VAEs require fewer computational resources than all the DGMs, converge faster, and one can have more control over the conditional parameters. They could be adopted for controllable sampling by either modifying the loss function to enforce latent variables to be correlated with properties of interest or to feed the conditional information to different parts of the model. VAEs can also be incorporated to compress data in hybrid DGMs.
However, VAEs face a few challenges~\cite{kingma_introduction_2019,tomczak_deep_2022}. A potential issue is the ``posterior collapse''~\cite{bowman_generating_2016}, which occurs when a powerful decoder treats the latent variable $z$ as mere noise, leading to the regularization term being minimized for priors like the standard Gaussian. 
Another problem, known as the ``hole problem'' \cite{rezende_taming_2018}, arises from a mismatch between the aggregated posterior and the prior. If there are regions where the prior assigns high probability, but the aggregated posterior assigns low, sampling from these regions can result in low-quality output. 
Moreover, minimizing the KL divergence between the approximated and true posterior distributions, can lead to a variance mismatch when these distributions are not perfectly aligned. This issue is known as ``blurriness problem''~\cite{kingma_introduction_2019} and is addressed by using more flexible inference and generative models to better capture the data distribution and improve sample fidelity.

\textbf{Normalizing Flows}. A normalizing flow~(NF) estimates the density of data $p(\mathcal{D})$ directly with an invertible and deterministic bijection between the latent variables and the data manifold via the change of variables formula. A typical instance of flow-based models takes the following form:

\begin{equation}
p(\mathcal{D}) = p(f(\mathcal{D}))\left|\det\left(\frac{\partial f^{-1}(\mathcal{\mathcal{D}})}{\partial \mathcal{D}}\right)\right|,
\end{equation}

where $p(\bm{z})$ is the density of the latent variables $\bm{z}$. 
The term $f^{-1}(\mathcal{D})$ is the inverse of the transformation $f$ that maps the latent variables to the data. 
The determinant of the Jacobian matrix  $\left| \det \left( \frac{\partial f^{-1}(\mathcal{D})}{\partial \mathcal{D}} \right) \right|$ accounts for the change in volume induced by this transformation, allowing for efficient and expressive density estimation.
Flow-based models are proper candidates for lossless compression~\cite{hoogeboom_integer_2019,berg_idf_2021}, and Approximate Bayesian Computation~(ABC)~\cite{papamakarios_sequential_2019} in simulation-based inference since they allow the calculation of the exact likelihood. 
However, the likelihood does not always correlate directly with sample quality~\cite{theis_note_2016}.
Conditional flow-based models could also be used to form a flexible family of variational posteriors~\cite{rezende_variational_2016,berg_sylvester_2019} in VAE-based models with which the lower bound to the log-likelihood function could be tighter.

The main problem with NF-based models is the computational overhead they create with high-dimensional data. NF models can struggle with very high-dimensional data spaces as the complexity of the necessary transformations grows exponentially. 
This makes them less suitable for simulating complex and high granular detector signatures like the planned CMS High Granularity Calorimeter~(HGCAL)~\cite{magnan_hgcal_2017} for HL-LHC~\cite{bruning_chapter_2020} and the Pixel Vertex Detector (PXD)~\cite{giakoustidis_status_2023} at Belle~II~\cite{abe_belle_2010}.

\textbf{Generative Adversarial Networks~(GAN).} 
A GAN~\cite{schmidhuber_making_1990,goodfellow_generative_2014} is an implicit density estimator~\cite{noauthor_statistics_nodate}, which learns to sample data points in a zero-sum game. 
GANs consist of two main components, namely, a generator $f_G$ for estimating the density implicitly and generating realistic data and a discriminator $f_D$ for distinguishing between the generated and real detector responses. This adversarial mechanism encourages the generator to produce synthetic data that the discriminator cannot distinguish from real data, improving the model's ability to capture the data distribution. Although GANs do not learn the density directly, they do it implicitly, where one can indeed incorporate it for simulation-based inference~\cite{ramesh_gatsbi_2022}, for example, in detector unfolding~\cite{datta_unfolding_2018,bellagente_how_2020}.
Formally, its training objective is a min--max game as follows:

\begin{equation}
	\min_{f_\text{G}} \max_{f_\text{D}} \, \mathcal{L}_\text{GAN}(f_\text{G}, f_\text{D}) = \mathbb{E}_{\mathcal{D} \sim p(\mathcal{D})}[\log f_\text{D}(\mathcal{D})] + \mathbb{E}_{\bm{z} \sim p(\bm{z})}[\log (1 - f_\text{D}(f_\text{G}(\bm{z})))].
\end{equation}

GAN-based models, by design, allow easy implementation of controllable sampling due to introducing a sample discriminator given the desired properties, though taming its convergence is very difficult. In general, training GANs is a highly brittle and painful task. It requires a significant amount of patience and hyperparameter tuning for domain-specific tasks, but when it generalizes well, it can produce remarkably high-resolution samples~\cite{sauer_stylegan-xl_2022,noauthor_ffhq_nodate} with high fidelity.
The adversarial training makes GANs suffer from mode collapse and vanishing gradients. That is why during the past few years many extensions and auxiliary methods have been introduced to mitigate these issues~\cite{salimans_improved_2016,kodali_convergence_2017,arjovsky_wasserstein_2017,bellemare_cramer_2017,miyato_spectral_2018,brock_large_2019,karras_style-based_2019,cont_tail-gan_2023}.
Nevertheless, GANs provide good control over the conditioning and enable representation learning~\cite{kang_contragan_2021,hashemi_ultra-high-resolution_2023}. Representation learning refers to deriving meaningful and often stronger latent features from raw data that can be used effectively in various downstream tasks such as classification or prediction. 
However, a specific issue with GANs is their evasive Nash equilibrium, a state where the discriminator and the generator cannot unilaterally improve their position by deviating from their current strategy. The models may continue to oscillate without settling into a stable state, or they may converge to a non-optimal solution. This makes the model selection non-trivial, since GANs have no straightforward way to select the best training epoch.

\textbf{Autoregressive Models}
Autoregressive Models~(ARMs)~\cite{bengio_neural_2000,oord_conditional_2016,chen_pixelsnail_2018,you_graphrnn_nodate} are based on the chain rule of probability and decompose a joint distribution over $N$ random variables. They are designed to generate data in a sequential manner. Specifically, these model factorize the generation process as a sequential step, which determines the next step action given an initial detector layer, hit, jet constituent, or shower. The general formulation of ARM models is as follows:

\begin{equation}
p(\mathcal{D}_{e}) = \prod_{i=1}^{N} p(\mathcal{D}_{e}^i\mid \mathcal{D}_{e}^1,\mathcal{D}_{e}^2,\cdots,\mathcal{D}_{e}^{i-1}),
\end{equation}

where it is possible to directly maximize the likelihood of the data by training a recurrent neural network to model $p(\mathcal{D}_{e}^i\mid \mathcal{D}_{e}^{1:i-1})$ by minimizing the negative log-likelihood,

\begin{equation}
    -\ln p(\mathcal{D}_{e}) = -\sum_{i} \ln p(\mathcal{D}_{e}^i\mid \mathcal{D}_{e}^1,\mathcal{D}_{e}^2,\cdots,\mathcal{D}_{e}^{i-1}).
\end{equation}

Although these models might utilize internal hidden states during the generation process, it is worth noting that these states are not considered latent variables in the traditional probabilistic sense. Since ARMs work like sequential generation, applying these models requires a pre-specified ordering defined by hit orders $\mathbf{H}_e$ or sensor/layer $\mathbf{L}_e$ orders in a detector signature triplet. Thus, they offer the advantage of generating highly correlated data, thanks to their sequential conditioning on previously generated elements. This enables intricate probabilistic relationships to be captured, offering the potential for more realistic simulations.
While autoregressive models are extremely powerful density estimators, sampling is inherently a sequential process and can be exceedingly slow on high-dimensional data. 
Additionally, data must be decomposed into a fixed ordering~(\textit{e.g.}, temporal); while the choice of ordering can be clear for detector layers in regular detectors, it is not obvious for irregular and complex topologies, which in turn can affect performance. Hashemi et al.~\cite{hashemi_deep_2024} tried to solve this problem with proper a positional embedding and inductive bias injection.

\textbf{Diffusion Models} Diffusion models or deep score-based generative models ~\cite{song_generative_2020,ho_denoising_2020,song_how_2021,kingma_variational_2023,ramesh_hierarchical_2022} are a class of generative processes inspired by non-equilibrium thermodynamics~\cite{sohl-dickstein_deep_2015} and PDEs~\cite{tzen_neural_2019}. Score matching is based on the idea of minimizing the difference between the derivatives of the data and the model’s log density functions.
Diffusion models contain two processes, the forward and the reverse diffusion process. The forward diffusion process constantly adds noise to the data sample using a noise schedule $\beta_t \in (0,1)$ controlling the step size, while the reverse diffusion process recreates the true data sample from a Gaussian noise input. 
In the language of hierarchical VAEs~\cite{kingma_improved_2016,sonderby_ladder_2016}, the bottom-up path~(\textit{i.e.}, the variational posteriors) can be a diffusion process, and the top-down path is parameterized by a reversed diffusion. Ergo, they provide only an approximation of the likelihood.
Formally, the forward diffusion process from step $(t-1)$ to $t$ is defined as a discrete Markov chain

\begin{align}
q(\mathcal{D}_t\mid \mathcal{D}_{t-1}) & = \mathcal{N}(\mathcal{D}_t;\,\sqrt{1-\beta_t}\mathcal{D}_{t-1},\,\beta_t\bm{I}), \\
q(\mathcal{D}_{1:T} \mid \mathcal{D}_0) & = \prod_{t=1}^T q(\mathcal{D}_{t} \mid \mathcal{D}_{t - 1}).
\end{align}

Note that the reverse diffusion process $q(\mathcal{D}_{t-1}|\mathcal{D}_t)$ will also be Gaussian if $\beta_t$ is small enough. However, as $q(\mathcal{D}_{t-1}|\mathcal{D}_t)$ is intractable, $p_{\theta}(\mathcal{D}_{t-1}|\mathcal{D}_t)$ is learned to approximate $q(\mathcal{D}_{t-1}|\mathcal{D}_t)$ as follows:
\begin{equation}
p_\theta(\mathcal{D}_{t-1}\mid \bm{x}_t) = \mathcal{N}(\mathcal{D}_{t-1};\,\bm{\mu}_\theta(\mathcal{D}_t, t),\, \bm{\Sigma}_\theta(\mathcal{D}_t, t)),
\end{equation}
 By optimizing a re-weighted variant of the ELBO, 
 
\begin{equation}
-\log p_{\theta}(\mathcal{D}_0) \leq \mathbb{E}_{q(\mathcal{D}_{1:T}\mid \mathcal{D}_0)}\left[\log\frac{q(\mathcal{D}_{1:T}\mid \mathcal{D}_0)}{p_\theta(\mathcal{D}_{0:T})}\right],
\label{eq:diffusion-vlb}
\end{equation}

where $p_\theta(\mathcal{D}_{0:T}) = p(\mathcal{D}_T) \prod_{t=1}^T p_\theta(\mathcal{D}_{t-1} \mid \mathcal{D}_t)$. The ﬁnal objective takes expectation over $q(x_0)$

\begin{equation}
     \mathcal{L} = \mathbb{E}_{q(\mathcal{D}_{0:T})}\left[\log\frac{q(\mathcal{D}_{1:T}\mid \mathcal{D}_0)}{p_\theta(\mathcal{D}_{0:T})}\right].
\end{equation}

Along with the interoperability of diffusion models and their remarkable results in natural images domain~\cite{ho_denoising_2020,kingma_variational_2023,ramesh_hierarchical_2022}, recent studies~\cite{2024arXiv240700783F} are showing the advancement of representation learning with them unlike flow-based models where there is no information bottleneck~(the latent space has the same dimension as the data space). Moreover, diffusion-based models can be adapted to approximate priors in VAEs~\cite{vahdat_score-based_2021,wehenkel_diffusion_2021}. 
A downside to consider when applying diffusion models to high-granularity detector signatures is the computational cost. The diffusion process involves multiple steps, each of which typically requires its own round of computation, making them somewhat slow samplers and computationally intensive.

\begin{table}[!htb]
    \centering
    \caption{A generic comparison between Deep Generative Models~(DGM) for simulating detector signatures}
    \begin{tabularx}{\textwidth}{XXXXXXX} 
        \toprule
        DGMs & Training & Likelihood Estimation & Granularity & Sampling Time & Compression & Representation Learning \\
        \midrule
        VAEs  & Stable  & Approximate & High  & Fast  & Lossy & \cmark  \\
        Flows  & Stable  & Exact  & Low  & Slow/Fast  & Lossless & \cmark  \\
        GANs  & Unstable  & Implicit  & Ultra-High  & Fast  & Lossy & \cmark \\
        ARMs  & Stable & Exact  & Low/Mid  & Slow  & Lossless & \xmark \\
        Diffusion  & Stable  & Approximate  & High  & Slow  & Lossy & \cmark \\
        \bottomrule
    \end{tabularx}
    \label{tab:compare}
\end{table}

\subsubsection{Sampling Strategies}
After learning a latent space~(or initial inputs in the case of ARMs) for representing the input data, the generative model samples from the learned~(or latent) distribution during the inference~(\textit{i.e.} sampling). 
The sampling strategies could be divided into two categories: \emph{random sampling} and \emph{controllable sampling}.
Random sampling simply draws latent samples from a simple prior distribution~(Normal Distribution), in which the model learns to approximate the distribution of the observed detector responses. This corresponds to sampling from the distribution $p(\mathbf{D})=p(\mathbf{L}, \mathbf{C}, \mathbf{H})$. 
The latter, along with random noise sampling, samples detector signatures with controls over the desired properties~(\textit{e.g.}, detector geometries, kinematic parameters, energies, luminosities, and beam parameters) as well. This corresponds to sampling from the conditional distribution $p(\mathbf{H} | \mathbf{L}, \mathbf{C})$ or $p(\mathbf{H},\mathbf{L} | \mathbf{C})$, depending on the situation. 
Therefore, random sampling is relatively trivial, while controllable sampling usually requires extra effort in algorithm design. 

Controllable sampling usually manipulates the randomly sampled $z \sim p(z)$ or the encoded vector $z \sim p(z,\mathbf{L},\mathbf{C})$ in the latent space to obtain a final representation vector $z$, which is later decoded to the detector response representation $\mathbf{H}$ with expected properties. 
There are three types of commonly used approaches:

\begin{itemize}
    \item \textbf{Latent Disentanglement Methods:} Disentangled latent methods~\cite{bengio_representation_2014,kim_disentangling_2019,mathieu_disentangling_2019,collins_exploration_2022} factorize the latent vector $z$ to $n$ parts, each of which focuses on one property $p_n$, encouraging the learned latent variables to be disentangled from each other. Therefore, varying one latent dimension $z_n$ of the latent vector $z$ will lead to property change in the generated detector responses.
    
    \item \textbf{Control Code Methods:} This method~\cite{sohn_learning_2015,mirza_conditional_2014,winkler_learning_2023} incorporates a conditional vector $c$~(discrete or continuous) that explicitly controls the property of generated detector elements. In this case, the final latent representation is usually a modulation of $z$, $\mathbf{L}$ and $\mathbf{C}$. This method is the most common conditional sampling in detector signature simulation due to its versatility.
    
    \item \textbf{Iterative Editing Methods:} Edit-based methods~\cite{hjelm_iterative_2018,grover_graphite_2019,cheng_controllable_2020,winterhalder_latent_2021} focus on iterative refinements and involve a multi-step process to modify the embedding of latent vector $z$ with conditional priors $c$ to achieve the desired characteristics in the generated detector response. Unlike disentangled sampling or control code methods, where the modification is done in a single pass, iterative editing typically uses optimization-based strategies~(\textit{e.g.}, Slot-Attention mechanism~\cite{locatello_object-centric_2020,di_bello_conditional_2022}) to iteratively update the embedding of the latent vector. This enables fine-grained control of the generated data.
    
\end{itemize}

\subsubsection{Generation Strategies}
The generator~(in some models this is done by the decoder) restores the latent vector back to the data manifold. Due to the sparse, often high dimensional, and unordered nature of detector signatures~(before fixing the representation with voxelization or pixelization), the resulting outcome of the generator / decoder faces challenges in accurately reconstructing the data or generating new signatures. 
This could lead to artifacts or inaccuracies in the generated detector responses, if not handled carefully.
Thus, existing works take three types of generation strategies for detector response generation: one-shot generation, sequential generation, and zero-shot generation.

\textbf{One-shot generation}. In one-shot generation~\cite{rezende_one-shot_2016,kipf_variational_2016,wang_graphgan_2017,de_cao_molgan_2022}, an event is generated in one single step. It is achieved by feeding the latent representations to neural networks to obtain the desired representation. In practice, various neural networks could be utilized, including two-dimensional~(2D) Locally Connected Networks~\cite{de_oliveira_learning_2017}, 2D and three-dimensional~(3D) Convolutional Neural Networks~(CNN), Graph Neural Networks~(GNN), Multi-layer Perceptron~(MLP), or any combination of these modules according to different types of detectors and representations to be generated. 
The advantage of one-shot generation is that it generates the whole event data in a single step. 

If the detector topology has a non-sequential ordering~(non-cylindrical/cubical topology) or irregular geometry~(where sensor sizes vary with the detector depth as well as within a single layer), it is not always feasible to treat the sensor information, $\mathbf{L}$ the same as the hit manifold $\mathbf{H}$. 
Consequently, using a grid-based~(homogeneous) representation to characterize the full detector signatures is not the most efficient method. 
Instead, a variable multi-set~(heterogeneous) representation is preferable.  
Detector signatures in multi-set representation can be related similarly to the Mesh representation of 3D objects, which consists of a collection of vertices~(like detector hits) and polygon faces~(like layer information). 
For example, the Pixel Vertex Detector~(PXD)~\cite{giakoustidis_status_2023} detector at Belle~II carries such a non-trivial topology where the detector is a Toroid with an octagon inner layer and a dodecagon outer layer. 
Or the ATLAS experiment~\cite{noauthor_atlas_nodate} electromagnetic calorimeter incorporates an irregular geometry of calorimeter segmentation. As a result, in the one-shot generation of multi-set representations of detector elements, it is still of paramount importance to incorporate suitable inductive biases to capture the detector geometry and topology.

\textbf{Sequential generation}. In contrast to one-shot generation, sequential generation~\cite{bacciu_edge-based_2020,liao_efcient_nodate,ingraham_generative_nodate} generates the detector responses consecutively in a few steps. As there is ordering naturally deﬁned for detector layers such as a calorimeter detector, sequential generation has to follow a certain sequential inductive bias for the generation. This is usually done by generating probabilistic sensor or layer features while sampling and feeding step-by-step the reconstructed detector response following a predeﬁned ordering. Despite slow sampling, sequential generative models enjoy the beneﬁt of auto-regressive and causal reasoning, which prompts a precise correlation modeling of the data. 
Therefore, it could be easily combined with constraints in each of the generation steps when the responses to be generated should obey certain restrictions. 
Using sequential sampling, when generating detector signatures with either high-granularity or long detector layers, the error will accumulate at each step, possibly resulting in discrepancies in the ﬁnal generated and observed detector signatures. 
Also, in the case of non-cylindrical detector topologies such as PXD at Belle~II, pure autoregressive reasoning could introduce a non-existing sequential bias into the generated data. In this case, one has to change the perspective from the single ``shower'' generation to the full ``event'' generation, where correlations between different directions~(various angles) become important. Note that while ARMs rely on sequential generation by construction, the sequential generation strategy is more general an can also be applied to other DGM architectures, for example to break down the problem into easier tasks. 

\textbf{Zero-shot generation}. Zero-shot detector response generation refers to the process of generating detector responses without any prior exposure to the specific representation or structure of the ``new'' target detector signatures~\cite{hashemi_deep_2024}. The new target detector signatures could belong to beam parameters beyond the training range, higher incident energies, higher luminosities, or detector responses for uninstalled sensors, as will be discussed in~\cref{sec: ood}.
The term \textit{Zero-Shot} comes from the concept of zero-shot learning~\cite{noauthor_deep_nodate}, which is when a model can recognize or generate outputs for new, unseen categories or tasks without any training examples. This is achieved by leveraging the model's pre-existing knowledge and generalization abilities, typically acquired during the pre-training phase on large datasets by incorporating symmetries and constraints directly into the generative model.
This strategy~\cite{liu_attribute_2020,li_graph_2023,chan_deep_2021,padmakumar_extrapolative_2023} leverages the latent space's ability to capture the essential characteristics of the data manifold, enabling the generation of plausible detector responses for regions and conditions beyond the current data at hand. The core idea that is introduced in this study is to extrapolate to the Out-Of-Distribution~(OOD) data by designing a model to be more flexible and adaptive, allowing it to accommodate detector geometries, representations, and conditions beyond the training data.

\subsubsection{Evaluation}
The evaluation of DGMs is an active area of research in which several quantitative and qualitative measures have been proposed so far~\cite{krause_caloflow_2023-1,hashemi_ultra-high-resolution_2023,kansal_evaluating_2023,acosta_comparison_2023,hashemi_deep_2024,das_how_2023}.
Such an evaluation can be done at different levels, each with it's own advantages and disadvantages. In addition to metrics assessing the quality of generated samples, one should also evaluate the time required to generate the samples, as there is usually a trade-off between these two aspects~\cite{xiao_tackling_2022}. While a detailed comparison of all DGMs presented here is not possible, a subset of them will be compared to each other as part of the CaloChallenge~\cite{michele_fast_2022} in a dedicated summary publication~\cite{CaloChallenge_writeup}. 

\textbf{Histogram or Marginal Distribution-based Methods~(signature-level)}. The comparison of samples~(or quantities derived from them) between the DGM and the reference dataset in histograms is a fast way to evaluate the quality of a DGM. However, this method only provides a necessary, not a sufficient condition, as it is just a one-dimensional projection of a higher-dimensional space and therefore blind to correlations between the features.

\textbf{Downstream Physics-based Methods}. From the physics perspective, we are interested in the difference between using samples from the DGMs vs.~using samples from the traditional simulation chain. Therefore, instead of comparing the full distributions, one only focuses on the derived observables one actually cares about. While this method of evaluation the most aligned to our physics goals, it comes at the drawback that the selection of observables is fixed at the point of evaluation. The inclusion of additional observables at a later stage might reveal a suboptimal DGM. Some works included such a methods in their studies, investigating various reconstructed parameters such as the tracking impact parameters~\cite{hashemi_ultra-high-resolution_2023} or shower reconstructed energies~\cite{buhmann_hadrons_2021}. 

\textbf{Neural network~(NN)-based Methods}. These methods use the power of neural networks to learn correlations in the high-dimensional data, offering methods of evaluation that are sensitive to the entire distribution.  

One way of using NN-based methods is to do the evaluation of DGMs as two sample test. One can use the Neyman-Pearson lemma~\cite{noauthor_ix_nodate} and train a neural classifier to distinguish samples from the DGM from samples of the reference data~\cite{lopez-paz_revisiting_2018}. If a powerful classifier is not able to distinguish the two sets, one concludes that the two underlying distributions are identical. This approach provides not only a single number~(Area Under the receiver operating characteristic Curve~(AUC)) indicating the quality of the DGM, but also offers a way of analyzing which regions of phase space have problems~(via the distribution of the classifier outputs~\cite{das_how_2023}). The classifier-based test~\cite{krause_caloflow_2023-1} gives the most powerful result on the two-sample test. However, while these results are sufficient for the comparison, they are not always necessary. For example, the disagreement between the DGM and the reference distribution could be in an irrelevant region of phase space.

Another way of incorporating the NN-based methods is the feature space-matching methods~\cite{borji_pros_2022} that capture more complex aspects of the data distributions, including the diversity, patterns, and higher-level attributes of the data. They rely on a pre-trained~(on the real data) multi-class classifier~(on the conditions $C$) with any backbone model~\cite{salimans_improved_2016,zietlow_demystifying_2021,chong_effectively_2020,parmar_aliased_2022}. Then, one can calculate a distance function between the embedding space~(feature space) of the backbone network of the generated and the real data. The lower this distance, the higher the fidelity of the generated data~\cite{betzalel_study_2022}. In detector signature simulation, these metrics have shown a very high sensitivity to small changes in the real data distribution's quality~\cite{hashemi_ultra-high-resolution_2023,kansal_evaluating_2023} and diversity~\cite{hashemi_deep_2024}.

\subsection{Representative Work and Datasets}
\label{sec:rep}
In this section, we discuss representative works for deep generative models in detector signature simulation with an emphasis on how they handle generation and sampling. For each model, we first go through the works that target detector response~(shower and track) generation, then we review models that aim to do event or jet simulation. 

Shower signatures correspond to datasets that encapsulate the complex dynamics of particle showers in calorimeters. These datasets often include the spatial distribution of energy deposits, the type of interacting particle, incident angle, and other features that distinguish electromagnetic from hadronic showers.
Depending on the calorimeter design and the simulation framework, the granularity can range from very fine to coarser bins that encapsulate multiple layers of the calorimeter.

Jets are collimated sprays of particles that are produced when high-energy quarks or gluons~(partons) undergo a hadronization process. Jet signatures contain not only the final state particles in a jet but also their kinematic properties and relationships, enabling a multidimensional view of jets. The granularity in jet signatures is usually more variable, often depending on the event's complexity. Kinematic properties like transverse momentum, azimuthal angle, and pseudo rapidity can be considered at various resolutions, depending on the level of detail required for the specific analysis.

Track signatures come from the direct detector hits or fired pixels. They contain the signals corresponding to charged particle interactions with the detector elements. They often have extremely high granularities, mimicking the detector's actual resolution, which can be on the scale of more than $\mathcal{O}(10^8)$ channels. 
Each hit usually contains information like position, time, and the deposited charge of the hit, offering a granular view suitable for the downstream track reconstruction algorithms. 

In general, based on granularity, one can categorize DGMs into \num{4} classes: \emph{low granularity} data with $\mathcal{O}(100)$ channels like jets, \emph{mid granularity} data with $\mathcal{O}(10^4)$ channels, \emph{high granularity} data with $\mathcal{O}(10^5)$ channels, and \emph{ultra-high granularity} data with up to $\mathcal{O}(10^7)$ input channels such as High Granularity Calorimeter~(HGCAL)~\cite{magnan_hgcal_2017} at CMS~\cite{noauthor_phase-2_2017} with roughly \num{6} million channels, the PXD at Belle~II~\cite{abe_belle_2010} with more than \num{7.5} million-pixel channels, or the future EPICAL-2~\cite{noauthor_results_nodate} ultra-high granularity electromagnetic calorimeter with \num{12.5} million-pixel channels at ALICE~\cite{aamodt_alice_2008}.
In~\cref{tab:top}, we provide the top \num{15} state-of-the-art (SOTA) generative models with the highest granularities ever studied. 

Historically, each newly introduced architecture was focused on a specific detector layout and hence worked with its own dataset. 
Such a situation is unfavorable for a large-scale and inter-experiment deployment of DGMs for detector simulation, as it is almost impossible to study improvements of new architectures over older ones. 
To alleviate this problem, a public data challenge called the \textit{Fast Calorimeter Simulation Challenge 2022 --- CaloChallenge}~\cite{michele_fast_2022} was created. 
It consists of simulated showers of increasing dimensionality, ranging from $\mathcal{O}(10^2)$ to $\mathcal{O}(10^5)$ dimensions. 
These datasets can serve as a benchmark for the existing and future DGMs for mid to high-granularity experiments and will help to improve our understanding of common struggles, advantages, and disadvantages, as well as the scaling behavior of different DGM approaches. 

In addition, there are other public datasets with different granularities coming from various experiments. 
The Electromagnetic Calorimeter Shower dataset~(CaloGAN)~\cite{nachman_electromagnetic_2017,paganini_calogan_2018} was the first public dataset on calorimeter showers corresponding to energy deposits from positrons, photons, and charged Pions. 
Diefenbacher et al.~\cite{noauthor_photon_nodate,diefenbacher_new_2023} also provided a public dataset~(ILD dataset) for high-resolution photon showers conditioned on the incident energy and angle. 
JetNet~\cite{kansal_jetnet_2023} is another important benchmarking contribution to jet datasets, represented as multi-sets containing gluon, top quark, and light quark jets. JetClass~\cite{qu_particle_2022,qu_jetclass_2022}, a more complex public jet dataset, contains 10 classes of jets carrying three categories of features, kinematics, particle identification, and trajectory displacement information. 
Hashemi~et al.~\cite{hashemi_ultra-high_nodate,hashemi_ultra-high-resolution_2023} also recently introduced a public dataset~(PXD dataset) for ultra-high granularity detector signatures with a non-trivial detector topology of PXD at Belle~II, conditioned on the sensor position, \textit{i.e.} radius and angle. 
This dataset acts as a benchmark for evaluating the DGMs geared towards real experiments with ultra-high granularity and provides insights into the scalability of current DGM approaches towards the HL-LHC era. 
To spur further research and development, frameworks such as COCOA~\cite{charkin-gorbulin_configurable_2023} and the Open Data Detector~\cite{gessinger-befurt_open_2023,noauthor_acts_nodate} have been introduced recently. 

\begin{table}[!htb]
    \centering
    \caption{Current top \num{15} SOTA models based on the granularity of the signatures they can generate. 
    This list does not provide a fair comparison for the surrogate models simulating jet signatures as they inherently carry rather low granularities.}
    \renewcommand{\arraystretch}{1.6}
    \begin{tabularx}{\textwidth}{XXXXXX} 
        \toprule
        Model & Algorithm & Representation & Conditioning & Experiment & \textbf{Granularity}~$\uparrow$\\
        \midrule
        
        IEA-GAN~\cite{hashemi_ultra-high-resolution_2023,hashemi_pixel_2021}  & GAN  & grid/set & sensor position~(radius and angle)& Belle~II PXD (2023,2021) & $40\times250\times768 = \mathbf{7,680,000} $ ch \\

        WGAN~\cite{srebre_generation_2020}  & GAN  & grid & random & Belle~II PXD (2019) & $40\times250\times768 = \mathbf{7,680,000} $ ch \\
        
        YonedaVAE~\cite{hashemi_deep_2024}  & VAE/ARM  & multi-set  & sensor position and Luminosity & Belle~II PXD (2023) & $\mathbf{110,000}$ points\\ 
        
        3DGAN ~\cite{khattak_fast_2021,belayneh_calorimetry_2020}  & GAN  & grid  & incident energy and angle & CLIC ECAL (2021, 2020) & $25\times51\times51 = \mathbf{65,025}$ ch\\
        
        BIB-AE~\cite{diefenbacher_new_2023}  & VAE/GAN/NF  & grid  & incident energy and angle & ILD ECAL (2023)& $30\times60\times30 = \mathbf{54,000}$ ch\\
        
        CaloScore~v2~\cite{mikuni_caloscore_2023}  & Diffusion & grid  & incident energy and time information & CaloChallenge D3~(2023)& $45\times50\times18= \mathbf{40,500}$ ch\\

        iCaloFlow~\cite{buckley_inductive_2023}  & NF/ARM & grid  & incident energy & CaloChallenge D3~(2023)& $45\times50\times18= \mathbf{40,500}$ ch\\
        
        CaloDiffusion~\cite{amram_denoising_2023}  & Diffusion & grid  & incident energy & CaloChallenge D3~(2023)& $45\times50\times18= \mathbf{40,500}$ ch\\

        CaloScore~\cite{mikuni_score-based_2022}  & Diffusion & grid  & incident energy and time information & CaloChallenge D3~(2022)& $45\times50\times18= \mathbf{40,500}$ ch\\
        
        BIB-AE~\cite{buhmann_hadrons_2021}  & VAE/GAN  & grid  & incident energy & ILD AHCal (2021)& $48\times25\times25 = \mathbf{30,000}$ ch\\

        BIB-AE~\cite{buhmann_getting_2021}  & VAE/GAN  & grid  & incident energy & ILD ECAL (2021)& $30\times30\times30 = \mathbf{27,000}$ ch\\

        WGAN~\cite{li_generative_2023} & GAN & grid  & energy and the deposit coordinates & EXO-200 (2023) & $74\times350 = \mathbf{25,900}$ ch\\
        CaloClouds~\cite{buhmann_caloclouds_2023,buhmann_caloclouds_2023-1}  & VAE/Diffusion & multi-set & incident energy and cardinality & ILD ECAL (2023) & $\mathbf{6,000}$ points\\
        
        NF-VAE~\cite{abhishek_variational_2019} & VAE/NF & grid  & random & Hyper-k IWCD (2019) & $19\times16\times40 = \mathbf{12,160}$ ch\\
        
        SuperCalo~\cite{pang_supercalo_2023}  & NF & grid & incident energy and geometry & CaloChallenge D2~(2023) & $45\times 16\times 9 = \mathbf{6,480} $ ch\\
        
        \bottomrule
    \end{tabularx}
    \label{tab:top}
\end{table}

\textbf{Variational Autoencoders}. 
The work of Abhishek et al.~\cite{abhishek_variational_2019} is historically the first application of VAEs in detector simulation. They utilized the VAE with a Normalizing Flows~\cite{rezende_variational_2016} learnable prior, to recover the true posterior distribution better and to improve sampling for Water Cherenkov detector~\cite{proto-collaboration_hyper-kamiokande_2018} simulation with grid-based data of size $19\times16\times40$. 
This is one of the pioneering works in VAE applications for shower simulation that unfortunately got ignored entirely by the community. 

In~\cite{ghosh_deep_2020}, the ATLAS collaboration conditioned the encoder and decoder directly in a vanilla VAE, on the energy of the incident particle to generate showers corresponding to a specific energy.

In~\cite{deja_end--end_2020}, the authors study the performance of the Sinkhorn Autoencoder~\cite{patrini_sinkhorn_2019} and leverage it to have a trainable prior approximation, namely a noise generator to encode and generate embeddings with the same distribution in the latent space. In order to overcome the mode collapse issue and promote diversity, they include additional regularisation on the autoencoder's latent space. Following~\cite{ayinde_regularizing_2019}, they compute a similarity matrix for the neural network’s weights according to the cosine similarity between its different layers to assess the diversity.

In the context of Cherenkov detectors, DeepRICH~\cite{fanelli_deeprich_2020}, designs a conditional latent space as a combination of CVAE~\cite{sohn_learning_2015} and infoVAE~\cite{zhao_infovae_2018}, where the latent variable $\sigma$ is determined using a Bayesian Optimization~\cite{snoek_practical_2012}. The control variables in this conditional VAE are the kinematic parameters of each particle learned by an auxiliary classifier over the encoded latent manifold as a regularization. They only consider the reconstruction of their dataset.

Moving to even higher-dimensional calorimeter setups, in~\cite{buhmann_decoding_2021,buhmann_getting_2021}, the authors utilize the BIB-AE~\cite{voloshynovskiy_information_2019} model, conditioned on the incident photon energy. Thus, the latent manifold is conditioned on the Energy. They introduce a post-processing module that relaxes the trade-off between the accuracy of the emulated hit energy spectrum and the reproduced shower shape. This module is an MLP-based network that fixes the hit energy spectrum resolution between the input and generated images. 
In the later efforts~\cite{buhmann_hadrons_2021} for hadronic showers, along with improving their model, inspired by~\cite{otten_event_2021}, the authors use a Kernel Density Estimator~(KDE)~\cite{parzen_estimation_1962} to fit the learned latent manifold for the inference time and use rejection sampling for the correct density estimation. 
In an updated version of their work, McKeown et al.~\cite{diefenbacher_new_2023} add multi-parameter conditioning over the BIB-AE model, now conditioning on both incident angles and energies. Due to the multi-parameter space of their conditioning, they used Normalizing flows for latent space sampling during inference. 

To address the sparsity of the datasets, GVAE~\cite{hariri_graph_2021} introduces a graph-based VAE architecture for learning the representation of collision events without any controllable sampling for emulation. 
Abhishek et al.~\cite{abhishek_calodvae_2022} incorporated a Discrete Variational Autoencoder~(DVAE) based model~\cite{rolfe_discrete_2017,vahdat_dvae_2018,khoshaman_gumbolt_2019} with hierarchical dependencies of latent variables and a Restricted Boltzmann Machine~(RBM)~\cite{montufar_restricted_2018} latent prior using block Gibbs sampling for generation of the calorimeter showers. They also tackle the sparsity of the showers with a learnable masking like~\cite{musella_fast_2018}. 
This solves the assignment problem with the sparsity, but it will be difficult to extend this to the simulation of high-granularity detector signatures.

Huang et al.~\cite{huang_fast_2023}, introduce an Autoencoder model for lossy compression of the data from Time Projection Chambers~(TPCs) in sPHENIX experiment~\cite{noauthor_bnl_nodate}. 
Their model, BCAE++, is an improvement over the original BCAE~(Bicephalous Convolutional Autoencoder)~\cite{huang_efficient_2021} and is designed to handle the sparse data, achieving a higher compression ratio and better reconstruction accuracy. It utilizes a specialized loss function that combines a segmentation decoder for voxel-wise bi-class classification and a regression decoder for reconstruction. This dual-decoder approach allows the model to effectively handle the sparse and irregular distribution of the TPC data. A novel aspect of this work is the application of half-precision mode in the network without too much loss in reconstruction accuracy.

Cresswell et al.~\cite{cresswell_caloman_2022}, develop a manifold hypothesis-inspired model~(density estimation)~\cite{noauthor_representation_nodate,brehmer_flows_2020,brown_verifying_2023} to make a dimensional reduction to the calorimeter data to speedup the inference-level sampling process. 
This method comprises a two-step approach. Initially, the manifold of calorimeter showers is learned using a generalized Autoencoder, which aids in constructing low-dimensional latent encodings. Following this, density estimation is performed to capture a probability density within the learned manifold, alleviating the dimensionality mismatch typically encountered in maximum-likelihood estimation. This structured procedure allows for a more efficient and practical simulation of calorimeter showers. A possible drawback of this approach might be the presumption that the manifold encompasses showers, which could potentially limit the ability to generalize beyond the training data~\cite{balestriero_learning_2021}.

To address the scalability issue of DGMs for ultra-high granularities with sparse representations, Hashemi et al.~\cite{hashemi_deep_2024}, introduced YonedaVAE, a novel multi-set generative model inspired by Category Theory~\cite{eilenberg_general_nodate,de_haan_natural_2020,dudzik_graph_2022}. 
For the first time, they do an Out-Of-Distribution~(OOD) detector simulation. Trained on low luminosity data of PXD at Belle~II with $\mathcal{O}(10^2)$ hit multiplicity, YonedaVAE generates high luminosity valid signatures with the correct intra-event correlation and ultra-high granularity of $110,000$ number of hits~(cardinality), without exposure to similar data during training.
YonedaVAE introduces a self-supervised set generator, capable of zero-shot creating sets and estimating the number of hits per sensor with variable ``inter-event'' and ``intra-event'' cardinality, facilitated by their Adaptive Top-q Sampling. 
Adaptive Top-q dynamically determines the multiplicity of points to be sampled based on the shape of the probability distribution for each event during inference. This adaptability is crucial for handling variable intra-event cardinality, especially when dealing with simulating the full detector with irregular detector geometries and hit patterns.
They showed that YonedaVAE could produce new detector signature point clouds with cardinalities well beyond the training data and achieve context extrapolation. This work can also be categorized as an ARM as it simulates each detector sensor and layer with a causal Transformer~\cite{radford_improving_2018} with a proper positional embedding.

For event generation, Otten et al.~\cite{otten_event_2021} introduce a method called buffering density information given the encoded events. They construct a prior by aggregating a subset of the encoded training data by saving all the parameters of the Gaussian distributions for all events in the training data to a file, which constitutes the buffer. At inference time, to increase the variance and avoid overfitting to the training data, they also sample from the buffered Gaussian distributions with a variance control factor. 

Orzari et al.~\cite{orzari_sparse_2021,touranakou_particle-based_2022} develop a VAE for generating constituents of hadronic jets represented by sets. They incorporate a permutation invariant loss, the Chamfer distance~\cite{fan_point_2016}, instead of the typical mean squared error~(MSE) as the reconstruction loss; however, their model does break the permutation equivariance using 2D convolution layers in the encoder. They regularize the model by constraining the $p_T$ and the invariant mass to follow the desired jet characteristics. The authors later in~\cite{tsan_particle_2021}, fix this issue by using Dynamic Graph Convolutional~(DGCNN)~\cite{wang_dynamic_2019} permutation equivariant layers.

Collins et al.~\cite{collins_exploration_2022,collins_machine-learning_2022}, using a ParticleFlow~\cite{komiske_energy_2019} $\beta$-VAE~\cite{higgins_beta-vae_2022}, find an interpretable and meaningful representation of the jets and their information complexity by analyzing the VAE's latent information. Moreover, they leverage $\beta$ from a fixed hyperparameter to an input of both the encoder and decoder networks.

\textbf{Generative Adversarial Networks}. 

The application of GAN-based generative models as implicit density estimators was embarked on by~\cite{de_oliveira_learning_2017} where they simulated 2D jet images for high energy W bosons and QCD jets as their conditional classes while introducing 2D locally connected layers~(LAGAN). 
CaloGAN~\cite{paganini_calogan_2018,de_oliveira_controlling_2018,paganini_accelerating_2018} employs the LAGAN layers to generate layer-wise two-dimensional images that were conditioned on the primary particle energy ranging uniformly from 1--100 GeV. 
Vallecorsa et al.~\cite{khattak_three_2018,vallecorsa_3d_2019,khattak_fast_2021} uses 3D Auxiliary
Classifier GAN~(ACGAN)~\cite{odena_conditional_2017}, to generate the calorimeter showers. In~\cite{belayneh_calorimetry_2020}, they add incident angle conditioning as well. Erdmann et al.~\cite{erdmann_generating_2018}, uses WGAN~\cite{arjovsky_wasserstein_2017} with continuous air shower energy conditioning using a constrainer network~(like ACGAN).
Musella et al.~\cite{musella_fast_2018}, for the generation of sparse hadronic jets using a U-net~\cite{ronneberger_u-net_2015} module for the generator, introduce a decision-making method by adding an additional channel to their output as a mask probability to decide if a pixel should be zero or not. 
Srebre et al.~\cite{srebre_generation_2020,} used the WGAN-gp~\cite{gulrajani_improved_2017} with random sampling to provide a proof of concept for the offline compression of the background hitmaps of ultra-high resolution Pixel Vertex Detector.

In~\cite{chekalina_generative_2019}, the authors do controlled sampling by conditioning a WGAN-gp model for LHCb calorimeter images. 
In~\cite{alonso-monsalve_image-based_2020}, they introduce an emulator-simulator setup that benefits from the Siamese Network~\cite{bromley_signature_nodate,chopra_learning_2005,koch_siamese_nodate}. They show that using their parameter-to-image grid 2-stage training pipeline, they can model complex functions. In the pre-training stage, the goal is to learn an emulator distribution that matches the Monte-Carlo simulator distribution using the Siamese network to learn the similarity of the simulated and emulated images. Then, at the next stage, a generator will be trained to learn to map the random noise to the parameter space. All these stages follow an adversarial training regime.

Hashemi et al.~\cite{hashemi_pixel_2021}, for the first time, generate the full ultra-high granularity PXD detector track hits with more than $7.5$ million pixel channels per event~\cite{hashemi_ultra-high_nodate} --- the highest spatial resolution detector simulation dataset ever analyzed with deep generative models in particle physics. 
Building upon their previous work, the authors later introduce the Intra-Event Aware GAN~(IEA-GAN) in~\cite{hashemi_ultra-high-resolution_2023}. 
As a fusion of the Transformer~\cite{vaswani_attention_2017} model and GANs with self-supervised learning and conditional contrastive loss, they introduce the Relational Reasoning Module to approximate the concept of an ``event'' in full detector sampling.
They do the conditioning on the position of the PXD sensors~(angles and radius) using methods from Deep Metric Learning~\cite{kaya_deep_2019}. 
Introducing self-supervised learning objectives in deep generative models, they show that they can generate fine-grained samples with large intra-event and small inter-event variations. Their evaluation encompasses a spectrum of results, extending from track-level~(signature-level) features to the downstream performance of Helix~(impact) parameter reconstruction.
It is noteworthy that calorimeter simulations typically focus on the simulation of particle showers from a single particle origin and in a small region of the calorimeter (\textit{i.e.} specifying one detector component of $\mathbf{L}$), which could indeed capture some aspects of inter-layer correlation within the scope of a localized area. However, IEA-GAN's approach extends this concept by considering the entire event with multiple-particle origins that encompass the full detector (\textit{i.e.} in all detector components $\mathbf{L}$) as a whole, where correlations among different sensors~(various angles and layers) become important within its readout window. 
Given the unique topology and geometry of the PXD as a highly granular tracking detector, this distinction is critical and allows for a more comprehensive simulation, capturing the complex interplay within an event across the entire detector rather than just the localized particle shower. 

Diefenbacher et al.~\cite{diefenbacher_dctrgan_2020} introduce a method for refining the precision of GANs using a post-processing re-weighting and tuning function based on~\cite{andreassen_neural_2020,badiali_efficiency_2020}. Kansal et al.~\cite{kansal_graph_2021,kansal_particle_2022} for the first time choose a more sparse representation of the detector data and introduce a graph GAN based on Neural Message Passing layers~\cite{gilmer_neural_2017}. Shirobokov et al.~\cite{shirobokov_black-box_2020} introduce a new approach that synergizes deep generative models and non-differentiable simulators. They show that one can both approximate the stochastic behavior of the simulator and enable direct gradient-based optimization of an objective by parameterizing the latent variable model with the relevant parameters of the simulator. Jaruskova et al.~\cite{jaruskova_ensemble_2023}, improve the calorimeter simulations over the lower energy depositions with AdaGAN-based~\cite{tolstikhin_adagan_2017} ensemble of GANs. 
Li et al.~\cite{li_generative_2023} use a WGAN model, conditioned on scintillation energy and the deposit coordinates, for simulating photodetector signals in the EXO-200 experiment's time projection chamber with grid-shaped data of size $74 \times 350$.

In~\cite{winterhalder_latent_2021}, Winterhalder et al. introduce the Latent Space Refinement~(LaSeR) protocol to enhance the precision and the topological obstruction of sampling by refining the predictions of a generator. In LaSeR, each generated sample is assigned a weight, which is then mapped to the corresponding latent space point. Rather than directly sampling from this weighted latent space, which could lead to biased results, the authors propose training a second generative model, the refiner, to transform the weighted latent space into an unweighted one. 

CaloShowerGAN~\cite{giannelli_caloshowergan_2023} employs Dataset 1 from the Fast Calorimeter Simulation Challenge 2022~\cite{michele_fast_2022} along with a well-structured training regime and data pre-processing to demonstrate revival of the classic CaloGAN~\cite{paganini_calogan_2018}. This study defines a new GAN-based baseline and addresses the void in the baseline comparisons for calorimeter shower simulation, superseding the CaloGAN. 

For the event/jet simulation, Hashemi et al.~\cite{hashemi_lhc_2019} use a GAN to directly emulate high-level features computed from the reconstructed $Z\rightarrow \mu \mu$ events. DijetGAN~\cite{di_sipio_dijetgan_2019} uses a simple GAN with random sampling for the simulation of QCD dijet events. 
Butter et al.~\cite{butter_how_2019}, do top pair generation with a modified MMD-GAN~\cite{li_mmd_2017} where the MMD kernel helps to describe on-shell resonances as well as tails of distributions, an improvement over~\cite{hashemi_lhc_2019}. 
They also study the statistical uncertainties and do ablation studies of the GAN approach to the event simulation using GANs.
Carrazza et al.~\cite{carrazza_lund_2019} employ CycleGAN~\cite{zhu_unpaired_2020} with the cycle-consistency loss to create mappings between two domains of Lund images~\cite{dreyer_lund_2018}, different categories of jets.
Farrell et al.~\cite{farrell_next_2019} apply GANs to generate full particle physics events conditioned on physics theory parameters.

Li et al.~\cite{li_polarization_2022}, introduce a style-based~\cite{karras_style-based_2019} conditional GAN~\cite{mirza_conditional_2014} to predict lepton decay angles in the rest frames of $W$ bosons in Vector Boson Scattering~(VBS) processes. Their approach addresses the challenge of missing neutrino information in the final state, which traditionally hampers the full determination of lepton angles. 
Alanazi et al.~\cite{alanazi_simulation_2021,velasco_cfat-gan_2022} develop a GAN-based model that does an importance sampling over the generated features that improve the sensitivity of the discriminator. 
Prieto et al.~\cite{bravo-prieto_style-based_2022} propose a style-based quantum GAN to generate events with a 3-qubit model. 
Howard et al.~\cite{howard_learning_2022} incorporate the Sliced-Wasserstein VAE~\cite{kolouri_sliced-wasserstein_2018} and the theory-based physics constraints in an unsupervised setting for event generation.
Epic-GAN~\cite{buhmann_epic-gan_2023}, by Buhmann et al., unveils an equivariant point cloud module for simulating particle jets as multi-sets, addressing the computational challenges with existing models. The framework, grounded on deep sets, showcases significant computational efficiency by avoiding pairwise information sharing between jet constituents. A notable merit of EPiC-GAN lies in its scalability to large particle multiplicities for jet data.

Käch et al. in~\cite{kach_attention_2023} propose a new GAN-based model that leverages attention mechanisms to generate particle cloud jets. 
The architecture is designed to handle only a variable inter-event number of particles. They utilize a Bert-like~\cite{devlin_bert_2019} aggregation mechanism that scales linearly with the particle multiplicity. As a result, they introduce a ``mean-field'' particle, which serves as an aggregation point for information from individual particles in the cloud. This mean-field particle is initialized by the sum of all particles and is used in a cross-attention mechanism~\cite{vaswani_attention_2017} to dynamically select which particles are important for distinguishing real from generated jets. This attention-based aggregation is particularly useful for handling large variances in individual particle energies. They achieve better results in jet simulation in comparison to the Epic-GAN~\cite{buhmann_epic-gan_2023} model.

Anderlini et al.~\cite{anderlini_generative_2023} introduce a distilled GAN from an ensemble of models~\cite{malinin_ensemble_2019} to reduce the variance for a maximally diverse set of models. In the inference time, they test the performance of their model in the Out-Of-Distribution~(OOD) regions of the phase space.

\textbf{Flow-based Models}.

For the first time, the CaloFlow~\cite{krause_caloflow_2023-1} applies the normalizing flow to the simplified calorimeter geometry of CaloGAN~\cite{paganini_calogan_2018} with \num{504} cells. 
Their model is a combination of Masked Autoencoder for Distribution Estimation~(MADE) blocks~\cite{germain_made_2015} and RQS transformations~\cite{durkan_neural_2019}. CaloFlow provides the additional benefit of tractable likelihoods with application to parameter inference for particle reconstruction. 
In CaloFlow~II~\cite{krause_caloflow_2023}, they incorporate knowledge distillation to transfer the probability
density of the stronger model~(teacher) to a much faster student model;  based on an Inverse Autoregressive Flow~(IAF)~\cite{kingma_improved_2016}. CaloFlow was then subsequently applied to \textit{CaloChallenge} dataset 1 in~\cite{krause_caloflow_2022}. 

SuperCalo~\cite{pang_supercalo_2023} embarks on a new perspective by employing a flow-based super-resolution model. 
This model targets fast coarse-grained calorimeter shower upscaling to high-dimensional fine-grained showers, providing a new point of view to Fast Simulation. The coarse voxel geometry is obtained from the full voxel geometry by grouping neighboring fine voxels to form coarse patches. The underpinning idea here is the utilization of conditional normalizing flows to recover the high resolution of voxel energies in calorimeters autoregressively patch-by-patch. 
However, using this method, recovering the global layer-by-layer correlation is still an issue.

For jet simulation, JetFlow~\cite{kach_jetflow_2022} employs Discrete Normalising Flows with multiple coupling layers. Each coupling layer divides input features into two sets, with one undergoing an identity transformation and the other a parameterized element-wise transformation. This configuration, with the addition of mass constraints, aids in modeling the generation of jets, showcasing a structured approach toward capturing more complicated correlations between the jet particles. Later, the authors~\cite{kach_point_2022} introduce a post-processing transformer encoder module, adversarially trained, to refine the output of the normalizing flow model.
Xu et al.~\cite{xu_generative_2023} develop a conditional Normalizing Flow based on~\cite{papamakarios_masked_2018} with the emphasis on modeling the correlation between the kinematic variables.

\textbf{Diffusion Models}.

CaloScore~\cite{mikuni_score-based_2022} marks the first application of a score-based generative model for detector simulation. They construct the score function using Conv3D U-nets models~\cite{ronneberger_u-net_2015}, conditioned on the normalized incident energy. 
In CaloScore~v2~\cite{mikuni_caloscore_2023}, the authors introduce an upgraded version of their previous model, which employs a single-shot diffusion model for detector simulation. The methodology involves the use of attention layers and progressive distillation techniques to reduce sampling times while retaining model performance.

CaloClouds by Buhmann et al.~\cite{buhmann_caloclouds_2023} as a novel combination of a VAE and diffusion model consists of four main components: the PointWise Net, EPiC Encoder~\cite{buhmann_epic-gan_2023}, Latent Flow, and Shower Flow. The PointWise Net serves as a permutation-invariant, diffusion-based point cloud generator. It is trained in parallel with the EPiC Encoder. The Latent Flow model is conditioned on energy and particle number and is trained simultaneously with the EPiC Encoder and the diffusion model. The Shower Flow is a separately trained normalizing flow model that estimates the number of points for a given incident energy and also serves for post-diffusion calibration. 
During inference, the CaloClouds architecture employs a three-step approach for sampling. Initially, a Shower Flow model is used to generate an appropriate number of points based on the incident energy. This Shower Flow also produces the total visible energy of the calorimeter point cloud and the number of points per layer for post-diffusion calibration. Once these variables are generated, the encoded latent space is produced using a conditional Latent Flow model. This Latent Flow is conditioned on the incident energy and the number of points generated by the Shower Flow. The final step involves using the PointWise Net to generate the point cloud through reverse diffusion. Currently, they could study point cloud signatures up to $6000$ cardinality.
The computational cost associated with the PointWise Net and the reverse diffusion process could be a bottleneck for high-granularity applications.

The authors later in~\cite{buhmann_caloclouds_2023-1}, with CaloClouds~II introduce several advancements over its predecessor, CaloClouds~\cite{buhmann_caloclouds_2023}. It adopts an EDM diffusion~\cite{karras_elucidating_2022} approach to speed up the sampling process by reducing diffusion iterations. For calibration, CaloClouds II shifts focus to per-layer energy calibration and X- and Y-direction center of gravity adjustments, moving away from the total energy calibration of CaloClouds, thereby improving longitudinal energy distribution fidelity. It also uses the consistency distillation introduced by Song et al.~\cite{song_consistency_2023}, enabling single-step data generation, which dramatically accelerates sampling speed.

Imani et al. ~\cite{imani_score-based_2023} focuses on the implementation and evaluation of a score-matching diffusion model for generating images from Liquid Argon Time Projection Chambers~(LArTPCs) signatures. The sampling involves the use of stochastic differential equations~(SDEs)~\cite{song_score-based_2021} to describe the diffusion process, with specific choices for drift and diffusion functions. 
In ~\cite{diefenbacher_refining_2023}, the authors introduce a novel proof of concept called Schrödinger bridge Quality Improvement via Refinement of Existing Lightweight Simulations~(SQuIRELS) to augment the quality of existing calorimeter simulation methods. The paper incorporates diffusion-based Schrödinger Bridge matching~\cite{shi_diffusion_2023,de_bortoli_diffusion_2023} to map between samples where the probability density is not explicitly known. 
The paper proposes a two-step process for refinement. The first step focuses on mapping the total energy sum of a given Fast Simulation to that of a high-fidelity simulation~(like Geant4). This model is a fully connected network consisting of three encoding networks. These networks are conditioned on the current time step and the incident particle energy. The second step involves a high-dimensional Schrödinger Bridge~\cite{noauthor_uber_1931} that refines the spatial distribution of the energy in the calorimeter. The paper employs forward and backward Gaussian transition kernels, utilizing neural networks to approximate forward and backward drift functions. 

Amram et al.~\cite{amram_denoising_2023} presents a denoising diffusion approach to simulating calorimeter showers, CaloDiffusion. Unlike CaloScore~\cite{mikuni_score-based_2022}, a score-matching network, they focus on the optimal denoising network~\cite{nichol_improved_2021}.
In order to solve the translation invariance problem of typical CNN models for calorimeter showers represented in a voxelized cylindrical
geometry, they use 3D cylindrical convolutions~\cite{zhu_cylindrical_2020}. Moreover, they introduce an interesting geometry latent mapping~(GLaM) module, which is able to map irregular detector geometries into a regular structure suitable for grid-based operations.

Acosta et al.~\cite{acosta_comparison_2023}, provides a comparison of point cloud versus image-based representation of calorimeter shower data with baseline diffusion Models~\cite{mikuni_caloscore_2023, mikuni_fast_2023}. For the image-based representation, they use $11\times11\times11$ dimensional grid, and for the point cloud version, they analyse the full $55\times55\times55$ dimensional sparse representation at the cell level with a maximum of \num{140} total number of hits. They found that the model adapted to the point cloud representation was better in generating shower data and showed a $3\times$ improvement in the sampling speed.

PC-JeDi~\cite{leigh_pc-jedi_2023} introduces new methodologies for generating jets as particle clouds. By leveraging transformers trained to reverse a diffusion process, PC-JeDi generates jets through an initial noise sampling for particle momenta, followed by denoising operations that capture complex underlying correlations with conditional generation. This approach facilitates the creation of jets with large transverse momentum from two distinct elementary particles. Later in~\cite{leigh_pc-droid_2023}, the authors achieve a faster and more accurate jet generation, using an EDM diffusion~\cite{karras_elucidating_2022} approach to speed up the sampling process and enhance the ability to perform the reverse diffusion process in fewer steps, cross-attention encoder~\cite{vaswani_attention_2017} as a faster and more memory efficient permutation equivariant network, along with the consistency distillation introduced by Song et al.~\cite{song_consistency_2023}. 

Mikuni et al.~\cite{mikuni_fast_2023} apply a Transformer-based diffusion model to particle jets~\cite{kansal_particle_2022} conditioned fully on the initial jet type, kinematics, and multiplicity. In order to increase the sampling process, they use the progressive distillation mechanism~\cite{salimans_progressive_2022} to transfer the knowledge of the Transformer-based teacher to an MLP-based student.

Butter et al.~\cite{butter_jet_2023} introduce two new diffusion models alongside an autoregressive transformer architecture for jet simulation. 
The first model, Denoising Diffusion Probabilistic Models~(DDPM)~\cite{ho_denoising_2020}, employs a time-dependent process that transforms a physics distribution into a Gaussian noise distribution. This is achieved by adding Gaussian noise in discrete steps, with each step having a specific variance. The reverse process then denoises this diffused data to recover the original distribution.
The second model also uses diffusion but with a continuous time evolution. 
The paper also introduces JetGPT, an autoregressive transformer architecture~\cite{noauthor_improving_nodate}, which aims to scale better with the phase space dimensionality. Bayesian versions of all three models are developed to control learning patterns and estimate uncertainties in density estimation.
For the DDPM, the sampling process makes the sampling process slower compared to classic generative networks. They illustrate the results first using two toy datasets, a two-dimensional linear ramp, and a Gaussian ring. Then, they applied all three networks to generate jet events, which allowed for comparative analysis, highlighting the advantages and disadvantages of the proposed architectures and the trade-offs between precision and computational efficiency.

\textbf{Autoregressive Models~(ARM)}.

Lu et al.~\cite{lu_sarm_2021} introduce a pioneering work, SARM~(Sparse Autoregressive Models for Scalable Generation of Sparse Images in Particle Physics), to model the joint distribution of the sparse jet images directly by breaking it down into a product of conditional distributions. Thus, each data point as a pixel is modeled autoregressively, conditionally dependent on the previous pixel in a certain reading pattern. This model takes sparseness into account by explicitly learning it with a tractable likelihood, providing a more stable and interpretable solution than the previous candidates for jet generation.

Liu et al.~\cite{liu_geometry-aware_2022,liu_generalizing_2023}, generate calorimeter responses autoregressively while taking into account variable detector sizes as a geometrical~(detector size) conditioning for OOD detector geometries. The authors employ a geometry-aware ARM that adapts its energy deposition based on the size and position of the calorimeter cells. The model is trained on a range of calorimeter geometries, including challenging transition regions where cell sizes abruptly change. 
The ARM framework consists of three Discrete modules, each trained separately for different layers of the calorimeter. They are based on MADE~\cite{germain_made_2015} and are designed to generate all desired parameters in a single pass, thus enabling faster training on GPU. 
The model takes into account not only the energy deposits but also the sizes of the cells, allowing it to generate energy distributions for different geometries adaptively. During training and inference, the model starts generating energy deposits from the central cell of the calorimeter, which typically receives the most energy deposition. The model then proceeds to generate energy deposits in the surrounding cells based on learned multinomial probability distributions in a spiral path.
This is particularly useful for handling non-uniform cell sizes and complex geometries.

Layer-to-Layer-Flows~\cite{diefenbacher_l2lflows_2023} introduce an autoregressive flow-based model for fast calorimeter emulation. 
The main idea of L2LFlows is to use CaloFlow with a causal inductive bias. Conditioned on the incident energy and a subset of preceding calorimeter layers for layer-to-layer associations, each layer of the calorimeter is learned by a separate flow. 
Inductive CaloFlow~\cite{buckley_inductive_2023} enhances L2LFlows, implementing a singular flow for generating calorimeter layer $i$ based on the shower shape of layer $i-1$, permitting an ``inductive'' learning of calorimeter showers while training not on entire events over all layers, but rather on pairs of layers. iCaloFlow employs in total three normalizing flows, one for the energy depositions per layer, one for the shower in the first layer, and one for the inductive step. 
Their teacher--student knowledge distillation framework also facilitates faster sampling while retaining high fidelity. However, the sampling time and scalability are still an issue with this approach.

For jet simulation, Di Bello et al.~\cite{di_bello_conditional_2022} introduce a modified Transformer model that is a combination of GNN as an encoding module and Slot-Attention~\cite{locatello_object-centric_2020} layers as a context-injecting layer to the probabilistic decoder which is a GRU-based~\cite{cho_learning_2014} model~\cite{cho_learning_2014}. Finke et al.~\cite{finke_learning_2023}, develops a masked Transformer Encoder-based model~\cite{vaswani_attention_2017}, based on TraDE (Transformers for Density Estimation)~\cite{fakoor_trade_2020} for discretized and ordered jet constituents.

\section{Applications}
In this section, we give an overview of the applications of DGMs in detector signature simulation. Speciﬁcally, we categorize them into four concrete branches: \emph{statistics amplification}, \emph{Amortised generation}, \emph{OOD simulation}, and \emph{anomaly detection}. Then, we illustrate their formulations in detector response generation and how different approaches and incorporation of inductive biases lead to success in various applications. 
``Inductive biases'' in DGMs refer to the assumptions the model makes to predict outputs for both in- and out-of-distribution inputs. These biases are inherent in the model's architecture and learning algorithm such as symmetries and adversarial robustness that guide the model's learning process~\cite{zhao_bias_2018,battaglia_relational_2018,zietlow_demystifying_2021,krippendorf_detecting_2020,dillon_symmetries_2021,barenboim_symmetry_2021,tombs_method_2022,desai_symmetrygan_2022,hashemi_ultra-high-resolution_2023,haochen_theoretical_2023}. 
In general, for DGMs in particle physics, the inductive biases could include distributional assumptions such as smoothness inductive bias, structural assumptions such as relational inductive bias or geometry awareness, and Physics-informed Assumptions such as Energy--Momentum conservation bias.

\subsection{Statistics Amplification}
As surrogate models are being used for detector signature emulation to do sample amplification~\cite{axelrod_sample_2019}, it is very important to quantify the statistical power of the generated dataset. Sample amplification is a procedure where there exists a map that takes a finite, initial subset of data and generates an extended one.
In this regime, DGMs work as classical parametric fits~\cite{carrazza_compressing_2021,chahrour_comparing_2022} to the training that also works as data augmentation methods. The question then would be to which extent they can interpolate. 
In other words, \emph{how can one quantify the diversity and uncertainty of DGMs?} This involves understanding the limitations and potential biases of the DGMs and possibly introducing methods to quantify the diversity or amplification factors~\cite{hao_data_2019,axelrod_statistical_2022}. 
For amplification of statistics, the smoothness assumption where the physics probability densities are smooth is very natural. In DGMs, the smoothness inductive bias is an assumption that similar inputs will produce similar outputs up to model interpolation. Regularization techniques in machine learning, which are used to encourage the model to learn smoother functions, are a common manifestation of the smoothness inductive bias.

The smoothness inductive bias is studied in~\cite{butter_amplifying_nodate,butter_ganplifying_2021,bieringer_calomplification_2022} for a simplified VAE-GAN~\cite{buhmann_getting_2021}.
Their study revealed that generated shower samples contain less information compared to a single real data point in the low sample size regime. When the number of generated shower samples increases significantly, the information contained within the generated sample set eventually levels off. As a result, they demonstrate the ability of DGMs to not only sample from implicitly defined distributions but also to leverage enhanced interpolation or fitting capabilities. Similar results are being achieved in~\cite{hashemi_deep_2024} from an Information Theoretic perspective while introducing a versatile diversity measure for detector signature simulation. 

To improve interpolation, it is crucial to consider how to diversify samples. 
In their work, Hashemi et al.~\cite{hashemi_ultra-high-resolution_2023} enhance the diversity and complexity~\cite{haochen_theoretical_2023,shwartz-ziv_compress_2023} of generated samples by introducing Self-Supervised Learning~\cite{balestriero_cookbook_2023} techniques that can be incorporated in any DGMs. 
In particular, they introduce a Uniformity loss that helps the model to avoid mode collapse and to generate more diverse samples by maintaining the ``uniformity of information'' inductive bias for the discriminator. 
Kansal et al.~\cite{kansal_evaluating_2023} study in detail the quantifying of the diversity of the generated samples by analyzing various metrics and measures for the jet point cloud generation task. They show that the message-passing GAN~\cite{kansal_particle_2022} performs better than set-transformer~\cite{lee_set_2019}.

\subsection{Amortised Generation}
For detector simulation, amortized generation generally refers to a strategy where a computationally expensive simulation process is replaced or approximated by a faster and more efficient surrogate model and goes beyond just pure enhancement of statistics. 
Detector effect simulations and parametrization can be computationally intensive due to a high space- and time-computation complexity. 
An ``amortized'' approach would involve training DGMs on a large set of data. Once trained, this model can generate new simulations much faster than the original simulation process. The term amortized in this context refers to the fact that the upfront cost of training the model is spread out, or ``amortized'', over the many simulations that the model generates. 

In an amortized generation, the sub-tasks are either ``Fast Simulation'' to overcome the high time complexity of detector simulation or ``Data Compression''~\cite{srebre_generation_2020,huang_efficient_2021,collins_machine-learning_2022,bengtsson_baler_2023,huang_fast_2023} to overcome the high space complexity of detector simulation. 
Based on the different types of data manifolds, various inductive biases are being incorporated. 
One can categorize different representations into two main classes, ``geometry-dependent'' and ``geometry-independent'' representations. 
Geometry-dependent approaches look at the detector signatures as grid-like structures. For example, one-dimensional fixed representations come with sequential local and translation invariance~\cite{mu_photon_2021,regadio_synthesis_2022}. 
Two-dimensional, 2D, representation come with local and translation-invariant inductive bias, which was studied for showers, jets, and tracks in depth for different detectors.
Then, in order to capture the detector's layer-by-layer association and correspondence, 3d grid-based models~\cite{khattak_three_2018,vallecorsa_3d_2019,belayneh_calorimetry_2020,khattak_fast_2021,mikuni_score-based_2022} are studied. However, as a consequence of translation invariance, it has the drawback of stationary assumption over the temporal/spatial features. 

Geometry-independent approaches, on the other hand, are more suitable for simulating detector signatures due to the inherent variable length of data and their heterogeneity and sparsity. 
For example, graph-based models~\cite{hariri_graph_2021,kansal_graph_2021,tsan_particle_2021,kansal_particle_2022,di_bello_conditional_2022}, besides the variable-length assumption and relational inductive bias, it assumes graph isomorphism inductive bias as well. This bias ensures that the model focuses on the structural information contained in the graph rather than the specific labeling of nodes.
Another geometry-independent approach is considering the set representation of detector signatures. 
The most important property with set-based models are the permutation equivariant encoders for jet, shower or track constituents~\cite{finke_learning_2023,buhmann_epic-gan_2023,leigh_pc-jedi_2023,mikuni_fast_2023,buhmann_caloclouds_2023,hashemi_deep_2024} or sensors~\cite{hashemi_ultra-high-resolution_2023}, and possibly permutation invariant loss functions~\cite{tsan_particle_2021,di_bello_conditional_2022,kansal_evaluating_2023}. 
For normalizing flows, equivariance under permutation group action is more non-trivial. 
Proved by Köhler et al.~\cite{kohler_equivariant_2020}, given a Flow-based model $F$ such that the set creation yields an exchangeable distribution, the update is permutation equivariant and invertible, and $p_{\theta}$ denotes the model likelihood, then $(F, -\log p_{\theta})$ is permutation equivariant.

While maintaining the variable-length assumption, dropping the permutation equivariance would correspond to a sequential inductive bias of detector responses either layer-by-layer~\cite{hashemi_deep_2024,diefenbacher_l2lflows_2023}, or hit-by-hit~\cite{lu_sarm_2021,liu_geometry-aware_2022,liu_generalizing_2023}.
While autoregressive models show a better predictive capability~\cite{butter_jet_2023}, the sequential nature of these models can be a disadvantage when it comes to high-dimensional computation, as the sampling is rather slow. 

\subsection{Out-Of-Distribution~(OOD) Generation}
\label{sec: ood}
DGMs for OOD and zero-shot learning is an exciting area of research that holds significant potential across various fields~\cite{anishchenko_novo_2021,freschlin_machine_2022,yeh_novo_2023,gainza_novo_2023,rajak_autonomous_2021,ravuri_skilful_2021,madani_large_2023,li_zero-knowledge_2023} including, but not limited to, drug Discovery, material design, and weather forecasting. 
Traditional methods for the synthetic generation of objects with enhanced or specific properties are often iterative and costly, requiring extensive manual work or heavy computational resources. 
 
In contrast, DGMs with zero-shot capability can deal with new scenarios that are not explicitly present in the training data, making them highly desirable in a wide range of problems. 
OOD generation of detector signatures is an emerging field in High Energy Physics~(HEP) as well~\cite{paganini_calogan_2018,di_sipio_dijetgan_2019,howard_learning_2022,anderlini_generative_2023,liu_generalizing_2023,hashemi_deep_2024}. 
For example, for real PXD data at Belle~II~\cite{abe_belle_2010}, an important challenge in working is its reliance on actual experiments for gathering the necessary random triggers. 
Consequently, real luminosity and beam-parameter-dependent PXD background data that go beyond current experimental limits are unavailable, leaving us reliant on computationally demanding simulations. This underscores the critical need for a surrogate model that can effectively generalize to OOD luminosity regions.
The current main challenge remains to be the optimization of DGMs, based on available information while avoiding overfitting, and the generalization to cases in which information is scarce or altogether absent, such as extrapolation to beam parameters, energies, luminosities, and geometries where there are no data. 

The first example of such OOD detector response generation is the CaloGAN paper~\cite{paganini_calogan_2018}, where they very briefly show they can generate showers beyond the training incident energy conditions. 
DijetGAN~\cite{di_sipio_dijetgan_2019}, also discusses extrapolation to new Beyond the Standard Model~(BSM)-dependent OOD regions of dijet invariant mass.
Anderlini et al.~\cite{anderlini_generative_2023} evaluates the uncertainty of GANs in new momentum regions through Background efficiency comparison. Their analysis shows that for Kaons, where the background efficiency does not decrease linearly in the OOD, high-momentum phase-space regions, their ensemble model fails to capture this behavior. 
Whereas, for Muons, due to the monotonic behavior of the background efficiency, the extrapolation shows better results.

A very important inductive bias for OOD generation is the variable-length assumption, especially in the context of detector signature simulation, due to the inherently variable nature of the data, characterized by a fluctuating number of detector responses in individual events. 
DGMs with a variable-length inductive bias have a promising capability to generate sets/sequences of differing lengths, contingent on the inherent complexity or dynamism of the event being generated. 
This equips the model with better adaptability when faced with novel and unseen scenarios.
Liu et al.~\cite{liu_geometry-aware_2022,liu_generalizing_2023}, by using an ARM approach~(length-independent), tries to extrapolate to unseen detector geometries where the extrapolation domain is calorimeter cell sizes. However, they show that the high granularity is a bottleneck to their approach.
Hashemi et al.~\cite{hashemi_deep_2024} address the challenge of OOD simulation for high granular and sparse events~(variable intra-event cardinality) by introducing YonedaVAE that can achieve both length and context extrapolation. 
Their model is trained on random trigger PXD background data from an early experiment of Belle~II with a peak recorded luminosity of \(1.42 \times 10^{34} \text{cm}^{-2}\text{s}^{-1}\) and a mean occupancy of 0.06\%~($\mathcal{O}(100)$ hit multiplicity). 
It was then tested on the previously unseen data from a later experiment of Belle~II, which had nearly double the peak luminosity --- \(2.68 \times 10^{34} \text{cm}^{-2}\text{s}^{-1}\) --- and a mean occupancy of 0.32\%~($\mathcal{O}(10^5)$ hit multiplicity). The length extrapolation here was when the model had to infer how an event with a higher~(than training) cardinality looks like. Context extrapolation means that the model has to produce the correct distribution of sparse cardinalities over all sensors, given only the maximum cardinality per event. 

\subsection{Anomaly Detection}
\label{sec:anomaly.detection}
Anomaly detection in high-energy physics is a large and growing field in itself~(see for example \cite{Kasieczka:2021xcg,Aarrestad:2021oeb} for reviews), but many of the proposed algorithms rely~(at least in part) on DGMs. In general, we identify three main ideas that are currently explored. 

The first one use the generative aspect of DGMs. The majority of these models are based on weakly supervised methods, especially the ``Classification WithOut LAbels''~(CWoLA) algorithm~\cite{Metodiev:2017vrx,Collins:2018epr,Collins:2019jip}. These methods require a signal-depleted background dataset, which can either be obtained in a data-driven way or from simulations by conditional generative models. Most of these anomaly detectors are based on Normalizing Flows~\cite{Hallin:2021wme,Raine:2022hht,Hallin:2022eoq,Mastandrea:2022vas,Golling:2022nkl,Sengupta:2023xqy,Golling:2023yjq,Finke:2023ltw,Bickendorf:2023nej,Bai:2023yyy} or diffusion models~\cite{Buhmann:2023acn,Sengupta:2023vtm}. There are also other approaches that use generative networks without the subsequent classifier step~\cite{Dillon:2019cqt,Dillon:2020quc,Fanelli:2022xwl}. 

The second group of anomaly detectors that use (parts of) DGMs is based on (V)AEs. This group uses the fact that the bottleneck architecture forces the encoder to learn a low-dimensional representation of ``usual'' data and anomalies can then be identified by a larger reconstruction error on the decoded output. It does therefore not rely on the generative nature of (V)AEs. This group constitutes the largest among all anomaly detectors that utilize generative architectures. In addition to only using autoencoders~\cite{Hajer:2018kqm,Heimel:2018mkt,Farina:2018fyg,Cerri:2018anq,Roy:2019jae,Blance:2019ibf,Amram:2020ykb,Cheng:2020dal,Pol:2020weg,vanBeekveld:2020txa,Bortolato:2021zic,Dillon:2021nxw,Finke:2021sdf,Atkinson:2021nlt,Govorkova:2021utb,Ostdiek:2021bem,Fraser:2021lxm,Herrero-Garcia:2021goa,Mikuni:2021nwn,Canelli:2021aps,Bradshaw:2022qev,Dillon:2022mkq,Golling:2023juz,Chekanov:2023uot,CMSECAL:2023fvz,Zhang:2023khv,Liu:2023djx}, some further use components of GANs~\cite{Knapp:2020dde,Vaslin:2023lig} or Normalizing Flows~\cite{Park:2020pak,Jawahar:2021vyu,Buss:2022lxw}.

The third group of anomaly detectors use DGMs that are at the same time density estimators. These can directly identify anomalous events by their small likelihood. Usually, these models only focus on the density estimation and anomaly detection aspect of the model~\cite{Nachman:2020lpy,Stein:2020rou,Caron:2021wmq,Verheyen:2022tov,Mikuni:2023tok,Das:2023bcj}, but as was pointed out in~\cite{Krause:2023uww}, they in principle have a double use: a single training of the DGM could be used for generating new samples and identifying anomalous events at no additional training costs, a promising new direction to optimize hardware resources.

\section{Outlook: Challenges and Opportunities}
Despite the remarkable advancements in deep generative models for efficient simulation of detector signatures, the field is far from reaching its full potential. 
In this section, we underscore various challenges and opportunities that have emerged from existing research and explore prospective directions for future investigations.

\textbf{Physics-Informed Generative Models.}
One of the most pressing challenges in the field of deep generative models for simulating detector signatures in particle physics is the incorporation of physics-based constraints or rules into the model architecture. While current methodologies often focus on data-driven approaches~(permutation equivariance is a special case), they sometimes overlook the underlying physics and symmetries that govern the data.
This could lead to models that are statistically accurate but physically implausible. 
Future work could focus on models that not only learn from data but also respect the governing symmetries~\cite{desai_symmetrygan_2022}, thereby ensuring both statistical and physical fidelity in the generated signatures.

\textbf{Precision and Uncertainty Quantification.} 
One of the less explored but critically important areas in the application of deep generative models for simulating high-granular detector signatures is the quantification of model uncertainty and the evaluation of precision~\cite{matchev_uncertainties_2022}. While these models can generate data that statistically resemble the target data, the question of how accurately they capture the underlying physical processes remains.
Traditional metrics~\cite{krause_caloflow_2023-1,kansal_evaluating_2023,hashemi_ultra-high-resolution_2023} like log-likelihood or NN-based metrics provide some measure of the model's performance but often fall short in capturing the nuances of physical validity and uncertainty. 
Moreover, the stochastic nature of these models introduces an inherent level of uncertainty in the predictions, which needs to be rigorously quantified to make the models truly useful. 
Future work could focus on the development of new evaluation metrics and techniques that not only assess the model's ability to replicate known phenomena but also its reliability in predicting new high-granular detector signatures. 
This could involve Bayesian approaches~\cite{butter_generative_2023,das_how_2023,heimel_precision-machine_2023}, ensemble methods~\cite{anderlini_generative_2023}, or even hybrid models~\cite{dherin_morse_2023} to provide error bars along with predictions.

\textbf{Real-Time Optimization.}
The increasing complexity of particle physics experiments, characterized by high event rates and pile-up conditions, poses significant challenges for real-time data analysis and decision-making for ultra-high granular detectors. Surrogate models, as discussed in this paper, provide very affordable solutions; however, optimizing their real-time performance is still a bottleneck. 
One promising avenue is the use of Federated Learning~(FL)~\cite{ratnayake_review_2023}, which allows for decentralized optimization across multiple nodes, thereby leveraging the computational resources of the entire experimental setup. FL enables models
to train on large real-time data sets and reduce biases associated with locally trained models.
Meta-learning techniques~\cite{naik_meta-neural_1992} are also strong candidates to enable models to quickly adapt to new experimental data or detector conditions, reducing the time required for retraining. 
Recently, the MetaHEP~\cite{salamani_metahep_2023} project embarked on this path.
Additionally, another promising avenue for future research lies in the direction of differentiable programming~(DP)~\cite{adelmann_new_2022,dorigo_toward_2022,aehle_progress_2023,kagan_branches_2023}. 
Through DP, software becomes differentiable through automatic differentiation~(AD)~\cite{baydin_automatic_2018}, enabling efficient gradient computation to understand the influence of input variations on output predictions. 
Such gradient-based insights could be invaluable for various downstream tasks. Utilizing DP frameworks would allow particle physics simulation tools to be integrated into machine learning pipelines in an end-to-end manner, thereby facilitating joint optimization for enhanced computational efficiency.

\textbf{Quantum Generative Models.} 
The advent of quantum computing offers a new avenue for tackling the computational challenges in particle physics simulations~\cite{di_meglio_quantum_2023}. Quantum Generative Models~(QGM) could potentially revolutionize the field by providing exponential speed-ups for certain types of problems. However, the practical implementation of QGMs for particle physics is still in its infancy~\cite{bravo-prieto_style-based_2022,rehm_full_2023,rousselot_generative_2023,hoque_caloqvae_2023}, and significant challenges related to error correction, qubit stability, and algorithmic design remain. 
Future research could focus on the development and validation of QGM architectures specifically tailored for detector signature simulations.

\textbf{Extrapolation Beyond Training Data.} 
One of the limitations of current generative models is their ability to generalize beyond the scope of the training data. 
Given the high-dimensional nature of HEP data, the inevitability of sparse datasets arises due to limited sample sizes, leading us into zones with inadequate training data. If the model demonstrates strong extrapolation capabilities, it does not just fill gaps in areas completely lacking data; it also provides a more reliable interpretation of regions where data is minimal. 
This is particularly crucial in particle physics, where there is a need to explore uncharted territories of parameter space~\cite{howard_learning_2022}, kinematic regions~\cite{butter_machine_2023,anderlini_generative_2023}, luminosities~\cite{hashemi_deep_2024} or detector geometries~\cite{liu_generalizing_2023}.  
Developing models capable of reliable extrapolation is an open challenge and involves techniques like uncertainty quantification and the incorporation of prior physical knowledge and inductive bias into the model.

\textbf{Scalability and Ultra-High Granularity Challenges.} 
The next generation of particle detectors will feature ultra-high granularities, as we defined in~\cref{sec:rep}, leading to an explosion in the dimensionality of the data~\cite{apostolakis_detector_2022}, for instance, the high Granularity Calorimeter~(HGCAL)~\cite{magnan_hgcal_2017} at CMS~\cite{noauthor_phase-2_2017} with roughly \num{6} million channels, the ultra-high resolution PXD at Belle~II~\cite{abe_belle_2010} with more than \num{7.5} million-pixel channels, or the EPICAL-2~\cite{noauthor_results_nodate} electromagnetic calorimeter with \num{12.5} million-pixel channels.
Except for some efforts~\cite{hashemi_ultra-high_nodate,hashemi_deep_2024}, current generative models are far from handling such ultra-high-dimensional data efficiently. Even these aforementioned efforts have not delved deeply into the issue of uncertainty quantification.
This gap underscores the urgent need for future work that not only addresses the development of scalable algorithms and architectures that can handle ultra-high granularity detectors with irregular geometries but also rigorously quantifies the uncertainties to ensure the physical validity and reliability of the generated detector signatures.

\section{Conclusion}
In this review, we offered an exhaustive and unified overview of deep generative models for the efficient simulation of detector signatures in diverse experimental settings. As a result, we introduced a cohesive framework to articulate the task and provide an up-to-date and comprehensive taxonomy of algorithms. Subsequently, we conducted an in-depth survey of existing techniques and models, elaborating on their key characteristics. We then turned our attention to three primary application domains where deep generative models are making significant contributions to the simulation of detector signatures. 
In the end, we spotlighted the prevailing challenges in current research and offered a panoramic perspective on the future avenues for advancing deep generative models in the realm of efficient detector signature simulation.

\section*{Acknowledgments}
This research was a chapter of Baran~(Hosein) Hashemi's PhD thesis~\cite{hashemi_deep_2024}, supported by the collaborative project IDT-UM~(Innovative Digitale Technologien zur Erforschung von Universum und Materie) funded by the German Federal Ministry of Education and Research~(BMBF) and the Deutsche Forschungsgemeinschaft under Germany’s Excellence Strategy – EXC 2094 ``ORIGINS'' – 390783311. 
We thank Anna Zaborowska for discussions. 
The authors would like to express their gratitude to the HEP ML Living Review~\cite{hep_ml_community_living_nodate}, which provided invaluable assistance in compiling and locating an extensive body of research.

\printbibliography

\end{document}